\documentclass[12pt]{article}
\pdfoutput=1

\usepackage{cite}

\usepackage[nointlimits,reqno]{amsmath}
\usepackage{amssymb}
\usepackage{amsmath}
\usepackage{graphicx}
\usepackage{tabularx}
\usepackage{multirow}
\usepackage{epstopdf}
\usepackage{slashed}
\usepackage{verbatim}
\usepackage[section]{placeins}
\usepackage[space]{grffile}
\numberwithin{equation}{section} 

\usepackage[pdftex,
 pdfproducer=TeXShop,
 pdfcreator=pdflatex]{hyperref}

\usepackage{xspace}
\usepackage{luty}

\begin{document}

\begin{titlepage}

\title{Effective WIMPs}

\author{Spencer Chang$^a$, Ralph Edezhath$^b$, 
Jeffrey Hutchinson$^b$, and Markus Luty$^b$}
   
\address{$^a$Institute of Theoretical Science, University of Oregon\\ 
Eugene, Oregon  97403}

\address{$^b$Physics Department, University of California, Davis\\
Davis, California 95616}

\begin{abstract}
The `WIMP miracle' for the relic abundance of thermal dark matter
motivates weak scale dark matter with renormalizable
couplings to standard model particles.
We study minimal models with such couplings that explain dark matter as 
a thermal relic.
The models contain a singlet dark matter particle with cubic renormalizable 
couplings between standard model particles and `partner' particles with the 
same gauge quantum numbers as the standard model particle.
The dark matter has spin $0$, $\frac 12$, or $1$, and may or may not be
its own antiparticle.
Each model has 3 parameters: the masses of the dark matter and standard model
partners, and the cubic coupling.
Requiring the correct relic abundance gives
a 2-dimensional parameter space where
collider and direct detection constraints can be directly compared.
We focus on the case of dark matter interactions with colored 
particles.
We find that collider and direct detection searches are remarkably
complementary for these models.
Direct detection limits for the cases where the dark matter is not
its own antiparticle require dark matter masses to be in the multi-TeV
range, where they are extremely difficult to probe in collider experiments.
The models where dark matter is its own antiparticle are strongly constrained
by collider searches for monojet and $\text{jets} + \text{MET}$ signals.
These models are constrained by direct detection mainly 
near the limit where the dark matter and partner masses are nearly
degenerate, where collider searches become more difficult.
\end{abstract}

\end{titlepage}

\section{Introduction}
\label{sec:intro}
The existence of dark matter is the strongest evidence we have for
physics beyond the standard model, and
it is a striking fact that a neutral particle with a mass at the
weak scale with renormalizable couplings to standard model particles
has a thermal relic density of order the observed value.
This `WIMP miracle' is a strong hint that motivates searches for 
direct detection of dark matter as well as dark matter production at colliders.
In the coming years, dark matter direct and indirect detection experiments
will reach new frontiers of sensitivity, and the LHC will begin operation at
14~TeV after a very successful 8~TeV run.
These experiments will provide a crucial test of these ideas,
and there is good reason to expect spectacular discoveries.

Naturalness also gives a motivation for new physics beyond the standard model
that has been very influential in particle physics.
The best-motivated and most successful framework for physics beyond the
standard model is supersymmetry (SUSY).
Among the attractive features of SUSY is that it contains a natural WIMP
candidate, the lightest supersymmetric particle (LSP).
Because of this, most of the work on the connection between direct detection
and collider searches for dark matter have focused on SUSY.
However, there is currently no signal for SUSY at the LHC, and minimal versions
of SUSY must be fine-tuned to accommodate the observed Higgs mass and the null
results of SUSY searches.
Also, LSP dark matter is only viable for special regions of parameter space:
either near maximal mixing, co-annihilation, or resonant annihilation.
All of these mechanisms require special relations between unrelated parameters.
More generally, the absence of any signal for physics beyond the standard
model at the LHC has led many to question whether naturalness is in fact
realized in nature, with or without SUSY. 

These considerations motivate a more phenomenological approach to dark matter,
one which assumes only the minimal extension of the standard model
required to account for dark matter.
One such approach that received wide attention is that of
`effective dark matter' 
\cite{Beltran:2008xg, Fan:2010gt, Fitzpatrick:2012ix,
Beltran:2010ww, Goodman:2010yf, Bai:2010hh, Goodman:2010ku, Aaltonen:2012jb, 
ATLAS-CONF-2012-147, CMS-PAS-EXO-12-048}.
In this approach, one assumes that the only new degrees of freedom relevant
for dark matter phenomenology are the dark matter particles themselves.
The only allowed interactions between the dark matter particles and the
standard model particles are non-renormalizable interactions of the form
\beq
\scr{L} \sim \frac{1}{M^n} | \text{SM} |^2 | \text{DM} |^2
\eeq
where SM and DM denote standard model and dark matter fields, respectively,
and $M$ is a mass scale that parameterizes the strength of the interaction.
The important point that these same operators parameterize direct detection
and monojet signals at colliders was made in 
\Refs{Goodman:2010yf, Bai:2010hh, Goodman:2010ku}.
For other work related to this approach, see
{\it e.g.}~\Refs{Birkedal:2004xn, Feng:2005gj, Cao:2009uw,  
Cheung:2012gi, Dreiner:2013vla}.  
This approach has many attractive features, but also several drawbacks.
First, the `WIMP miracle' that motivates weak scale dark matter is not
built in.
Also, collider bounds on higher dimension operators typically probe scales
$M$ of order the energy of the collisions, so the UV completion of the
operator becomes relevant for the collider phenomenology.

In this paper, we propose a different phenomenological approach to WIMP dark
matter that addresses these issues.
Motivated by the `WIMP miracle,'
we assume that dark matter has renormalizable interactions with standard
model fields.
We assume that the dark matter is a standard model gauge singlet.
The alternative is that the dark matter is the neutral component of an
electroweak multiplet, but this is highly constrained by direct detection
experiments.
Such models are viable models of dark matter only if the multiplet has 
$Y = 0$  and the dark matter mass is in the TeV range \cite{Cirelli:2005uq}.
For singlet dark matter, the only renormalizable couplings to standard 
model particles are quartic couplings of the Higgs boson
to scalar dark matter \cite{Silveira:1985rk,McDonald:1993ex, Burgess:2000yq, Cline:2013gha}.
This model has been extensively studied, and we only comment on it
briefly below.
Any other model with WIMP dark matter must contain additional degrees
of freedom.
We therefore consider cubic couplings of the form
\beq\eql{cubic}
\De \scr{L} \sim \la (\text{SM})  ( \widetilde{\text{SM}}) (\text{DM}),
\eeq
where $\la$ is a dimensionless coupling.
Here $\widetilde{\text{SM}}$ is an additional field with the same
gauge quantum  numbers as the standard model
field, so we call it a `partner' field.
This interaction is invariant under a $Z_2$ symmetry under which
the dark matter and partner fields are odd,
and therefore preserves the stability of the dark matter particle
as long as the dark matter particle is lighter than the
partner particle.

We focus on the case of interaction with colored standard model
particles, since this is the case of most relevance to both
the LHC and direct detection experiments.
This means that the partner fields are colored, and can therefore
be studied at the LHC.
We do not consider the case where the colored standard model particle
is the gluon because renormalizable interactions of this kind require
embedding $SU(3)_C$ into a larger gauge group at the weak scale,
and therefore require significantly more structure.
We also assume that the interaction is invariant under the 
electroweak gauge group.%
\footnote{Interactions with dimensionless couplings
that violate electroweak gauge symmetry can 
arise by integrating out TeV scale particles whose masses break
electroweak symmetry due to couplings to the Higgs field.
This requires additional structure, so we do not consider it
on the grounds of minimality.}
This type of interaction is familiar from SUSY,
where the partners are the superpartners,
but we see that it has a direct phenomenological motivation.

Interactions of the form \Eq{cubic} have
potentially serious problems with flavor physics,
a fact that is also familiar from SUSY.
For a generic flavor structure, loops involving virtual dark matter
and partner fields with weak-scale masses
will give rise to flavor-changing processes in
conflict with experiment.
The simplest solution to these constraints is to assume that the
couplings and masses of the partner fields are approximately 
flavor-independent.
The flavor constraints are most stringent for couplings to the first
two generations of quarks, so we will consider three cases:
$(i)$ 3 generations of quark partners with approximately equal couplings
and masses; 
$(ii)$ 2 generations of quark partners coupling to the first 2 generations
of quarks with approximately equal couplings and masses;
$(iii)$ a single quark partner coupling to third generation quarks.
To avoid proliferation of similar cases,
we consider only the case of coupling to left-handed quark doublets.%
\footnote{One case that could be significantly different is if the dark
matter couples only to the right-handed top quark.
However, the collider constraints come from top squark searches, which are similar in strength
to bottom squark searches.  The direct detection constraints are also qualitatively similar to the 
constraints for dark matter coupling to third generation left-handed quarks.  Thus, the limits on that case will be similar to our models coupling to the left-handed third generation quarks.}
From now on, we denote the dark matter particle by $\chi$, the left-handed
quark fields by $q$, and the quark partners by $Q$.
We consider dark matter with spin $0$, $\frac 12$,
or $1$, and the case where the dark matter is and is not its own antiparticle.
Strictly speaking, the case of vector dark matter falls outside
our minimal classification, since renormalizable interactions
of gauge fields will not connect quark fields and partner fields.
Generating such an interaction requires an extension of the electroweak
gauge group as well as additional Higgs fields.
We however include this case because it is phenomenologically similar
to the others, and the real vector case arises in 
universal extra dimension models \cite{Servant:2002aq, Cheng:2002ej}.
The quark partner fields are required to have masses that are invariant
under electroweak symmetry.
This means that the partners are complex scalars for fermion dark
matter, and Dirac fermions for bosonic dark matter.
We use 2-component spinor notation, so the Dirac mass for fermionic
quark partners is written $Q^c Q$.
The models are listed in Table~\ref{table:models}.

\begin{table}
\begin{center}
\begin{tabularx}{.85\textwidth}{|p{4cm}|p{4cm}|X|}
\hline
\multicolumn{2}{|c|}{Model Particles} & \multirow{2}{.2\textwidth}{\parbox{4cm}{\hspace{.5in} $\mathcal{L}_{\text{int}}$}}\\
\cline{1-2}
  Dark matter $\chi$ & Quark partner $Q$ &     \\ 
\hline
\hline
 Majorana fermion& Complex scalar&  $\la(\chi q)Q^*+\text{h.c.} $\\ 
\hline
Dirac fermion & Complex scalar & $\la(\chi q)Q^*+\text{h.c.}$ \\
\hline
Real scalar & Dirac fermion&  $\la (Q^c q)\chi+\text{h.c.}$  \\
\hline
Complex scalar & Dirac fermion & $\la (Q^c q)\chi+\text{h.c.}$   \\
\hline
Real vector & Dirac fermion & $\la \left(q^\dagger \bar{\si}^\mu Q\right)\chi_\mu+\text{h.c.}$\\
\hline
Complex vector & Dirac fermion & $\la \left(q^\dagger \bar{\si}^\mu Q\right)\chi_\mu+\text{h.c.}$\\
\hline
\end{tabularx}
\begin{minipage}{5.5in}
\caption{\small
Overview of the models considered in this paper.
Spinors are written in 2-component notation.
Here $q$ is the left-handed quark doublet of the standard model,
$Q$ is the quark partner field, and $\chi$ is the dark matter field.
\label{table:models}}
\end{minipage}
\end{center}
\end{table}

Because of the degeneracy assumed to avoid flavor constraints, each of these
models has 3 parameters: a dark matter mass $m_\chi$, a partner mass $m_Q$,
and a dimensionless coupling strength for the new cubic interaction $\la$.
Since the motivation for these models is the `WIMP miracle', we impose
the constraint that the thermal dark matter density is the observed value.
This results in a 2-dimensional parameter space that can be parameterized
by the masses $m_Q$ and $m_\chi$.
We have $m_Q > m_\chi$ by the assumption that the $\chi$ particle is stable.
As with the effective dark matter models, we can directly compare collider,
direct detections, and indirect detection constraints in a
2-dimensional parameter space.

\begin{figure}[t] 
\centering \hspace{.7in} \includegraphics[scale=1]{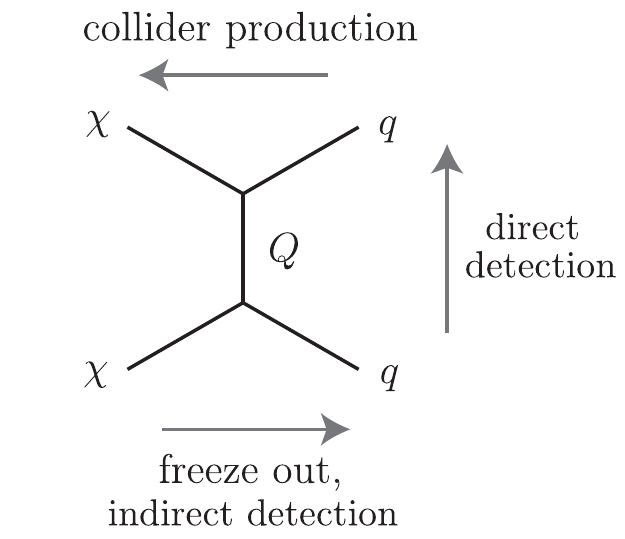}
\begin{minipage}{5.5in}
\caption{\small 
Feynman diagram contributing to dark matter
freeze-out, direct and indirect dark matter detection,
and collider production of dark matter.
\label{fig:basicint}}
\end{minipage}
\end{figure}

\begin{figure}[t] 
\centering \includegraphics[scale=0.6]{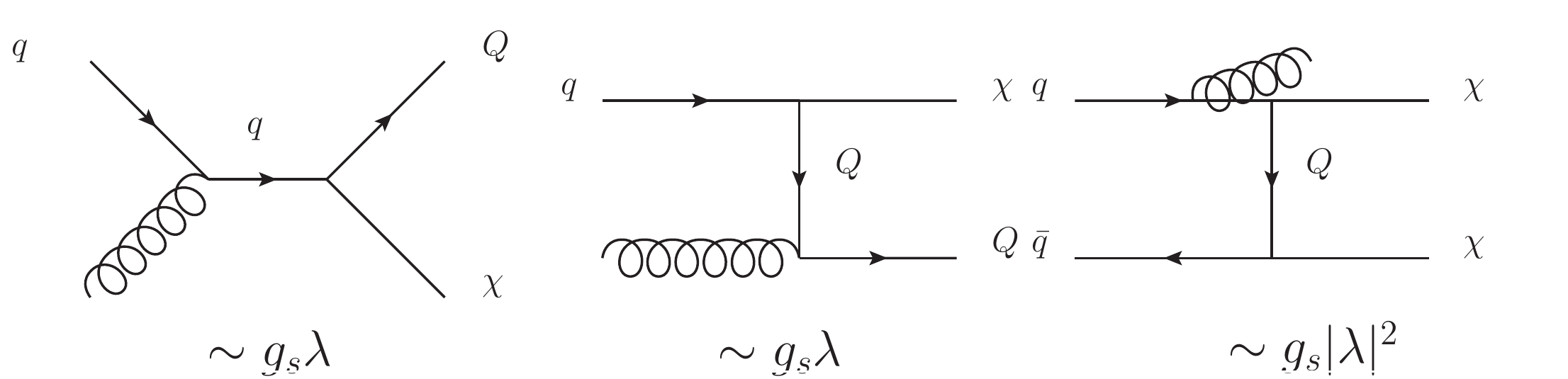}
\begin{minipage}{5.5in}
\caption{\small 
Feynman diagrams contributing to monojet signals at a hadron
collider.
\label{fig:monojet}}
\end{minipage}
\end{figure}

\begin{figure}[t] 
\centering \includegraphics[scale=0.6]{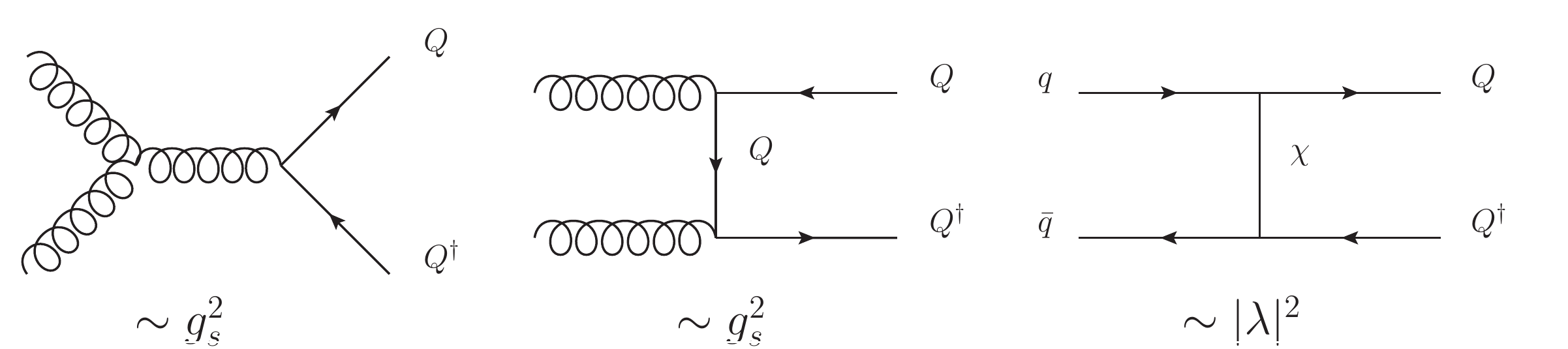}
\begin{minipage}{5.5in}
\caption{\small
Feynman Diagrams contributing to jets plus missing energy signals
at a hadron collider.  For scalar quark partners $Q$, there is an additional diagram involving the gluon-$Q$ quartic interaction that is not shown. 
\label{fig:jetsMET}}
\end{minipage}
\end{figure}

The annihilation of dark matter in the early universe,
indirect detection of dark matter,
and direct detection of dark matter are all dominated by the
exchange of a partner particle, as shown in Fig.~\ref{fig:basicint}.
The same diagram also gives rise to dark matter production at
colliders, with an additional radiative particle required
to tag the final state.
This strongly motivates monojet searches at the LHC as a way to
search for dark matter.
In the present models, there are additional contributions to
monojet final states, as shown in Fig.~\ref{fig:monojet}.
In addition, there are jets plus missing energy signals from diagrams 
such as  Fig.~\ref{fig:jetsMET}.
These models therefore have a very rich phenomenology controlled by a
simple 2-dimensional parameter space.

These models can be used in a number of different ways.
First, we advocate that they should be taken seriously as 
phenomenologically-motivated models  of dark matter under 
the assumption that a small number of states is relevant.
Another point of view comes from the fact that these models 
are also the minimal ones that can explain an excess in collider searches for
jets plus missing energy, perhaps the most promising channel for the 
discovery of SUSY.
If a signal is seen in $\text{jets} + \text{MET}$,
it would immediately raise the question of whether WIMP dark matter is
being produced in these events.
In the context of the models we are considering,
the rate and kinematics of such a signal would point to a specific
region of the parameter space, which can be
additionally probed by both monojet searches and direct detection
experiments.
A confirmation of the model predictions is clearly interesting,
while ruling out the model tells us that additional states are required
if the missing energy is due to WIMP dark matter.
Finally, these models can be viewed as `simplified models'   
\cite{Alves:2011wf} that parameterize
the constraints of experiments in terms of a model with only the
ingredients relevant for the signal.
In this case, they provide a well-defined mapping between collider 
and astrophysical constraints on dark matter based on  
a well-defined set of physical assumptions.
From all of these points of view, 
we believe these models can provide insight into
the complementarity between these different approaches to testing the
WIMP hypothesis.

Our main conclusion is that collider and direct detection experiments
are remarkably complementary under the assumptions we are making.
Models where the dark matter is not its own antiparticle have unsuppressed
direct detection cross sections, and current direct detection limits
require the dark matter mass to be in the multi-TeV range,
where they are extremely difficult to probe with colliders.
Models where the dark matter is its own antiparticle have suppressed
direct detection cross section, and direct detection limits are strong
near the degenerate limit $m_\chi \simeq m_Q$.
In this regime, direct detection is enhanced by small energy denominators,
while collider searches become more difficult due to smaller missing
energy.
Away from the degenerate limit, LHC searches are very constraining.  
Requiring the correct relic abundance gives a large (but still perturbative)
value of the new cubic coupling, which means that production by $t$-channel
$\chi$ exchange is dominant in a large region of parameter space.
For monojets there is a substantial contribution due to associated production of $Q \chi$.
Using existing monojet and jets + MET searches, we find that jets plus missing energy searches generally provide
the strongest collider constraint.  Due to the new processes, there could be improvements.  Because the kinematics of $t$-channel exchange differs from that of
standard colored production, we advocate the use of simplified models
that include the $t$-channel contributions in LHC searches.  Similarly, modifications to monojet searches could have enhanced sensitivity since the monojet's $p_T$ has a broad peak due to the two body decay of $Q$.  

We end this introduction by commenting briefly on other
phenomenological models with renormalizable
interactions with dark matter that have been considered in the literature.
One is a `Higgs portal' \cite{Patt:2006fw} interaction of the form 
$S^2 H^\dagger H$ between a 
singlet dark matter scalar $S$ and the Higgs field $H$ 
\cite{Silveira:1985rk, McDonald:1993ex, Burgess:2000yq, Cline:2013gha}.
There are two regions of parameter space where this model gives a
viable dark matter phenomenology.
The first is  the resonant annihilation
region $m_S \simeq \frac 12 m_h$, which can be probed both at collider
and direct detection experiments,
and a region with $m_S \gsim 80\GeV$ that can be probed by
direct detection, but is difficult to probe at colliders because
$h \to SS$ is forbidden \cite{Djouadi:2012zc}.
Another renormalizable phenomenological model is obtained by assuming
that  the dark matter is
the electrically neutral component of an electroweak multiplet \cite{Cirelli:2005uq}.
In order to avoid being ruled out by direct detection by $Z$ exchange
these multiplets must have vanishing hypercharge.
To get the right thermal relic abundance the dark matter must be
at the TeV scale, and these models contain no colored particles,
so these models are very difficult to probe at colliders.
By contrast, the models considered here where the dark matter is its own antiparticle are viable for a wide range of parameters
that will be extensively probed in both collider searches and direct dark
matter searches.  

An earlier work with a similar approach to ours considered only Majorana dark matter  \cite{Garny:2012eb}.   
Our analysis goes beyond \Ref{Garny:2012eb} by considering all allowed spins of the dark matter and quark partners and by considering additional collider processes.   

This paper is organized as follows.
In \S\ref{sec:Constraints}, we discuss the constraints the models from the
relic density, and direct and indirect detection and collider bounds.
In \S\ref{sec:Results} we present our main results,
and our conclusions are presented in \S\ref{sec:Conclusions}.  
The appendices contain detailed formulas used to obtain our results.

\section{General Features \label{sec:Constraints}}
In this section we consider the general features of the constraints
on the models from dark matter relic abundance,
direct and indirect detection, and LHC searches.
We also discuss the effective operators that describe the interactions
in the limit $m_Q \gg m_\chi$, since these allow one to understand
many of the qualitative features of the models.  We also consider the suppression and enhancement effects when the dark matter is nearly degenerate with the quark partner.

\subsection{Relic Abundance}

The relic abundance of non-baryonic matter is very accurately determined
by cosmological constraints
to be $\Omega_{\chi} h^2 = 0.1199 \pm 0.0027$ \cite{Ade:2013lta}.  
We assume that the dark matter is entirely composed of the $\chi$ particle in our model.
Under the assumption that $\chi$ particles were in thermal equilibrium in
the early universe, its present relic density is determined by freeze-out
from the annihilation process $\chi\bar\chi \to q \bar{q}$ shown in 
Fig.~\ref{fig:basicint}.
The relic abundance is determined by the thermally averaged annihilation
cross section $\avg{\si(\chi\bar\chi \to \bar{q}q) v}$
at temperatures $T_f \sim m_\chi / 25$.
The dark matter velocity is then $v^2 \sim 0.1$, so we can expand
\beq
\si(\chi\bar\chi \to \bar{q}q) v = a + b v^2 + O(v^4).
\eeq
Approximate formulas for the relic density in terms of these
parameters are given in Appendix A.
The coefficients $a$ and $b$ represent $s$-wave
and $p$-wave contributions,
and can be computed in each model.
Formulas for these are given in Appendix B.
 
As explained in \S\ref{sec:effint} below, 
the $s$-wave coefficient $a$ is suppressed by $m_q^2 / m_\chi^2$
in the cases where the dark matter is a scalar or Majorana fermion.
This arises because the dark matter couples only to left-handed quarks in our model.
The $p$-wave coefficient $b$ is suppressed by $m_q^2 / m_\chi^2$ only in
the case of real scalar dark matter.
Often the quark mass suppression of the $s$-wave is more severe than the velocity suppression of the $p$-wave which leads to larger couplings that vary rapidly with $m_\chi$.  
If these models couple to the top quark, the coupling $\la$ required
to get the right relic abundance
drops sharply for $m_\chi > m_t$ 
as the annihilation to top quarks becomes a viable channel.  

These suppressions play a very important role in the phenomenology of these
models, so it is worth considering the question of how robust this structure is.
We have assumed that dark matter couples to left-handed quarks, but this is
not particularly motivated over the assumption that the dominant coupling is
to right-handed quarks.
As long as we assume that only one quark multiplet dominates the phenomenology,
there will be a similar chiral suppression, and the results will be
qualitatively the same.
We can avoid this suppression only if the quark partners for left- and right-handed
quark fields have similar mass, including a large Dirac mass mixing them.
Such a mass breaks electroweak symmetry, and therefore must arise from couplings
to the Higgs.
These couplings are constrained by precision electroweak measurements,
and also require significant additional structure compared to the models
we consider here.
We conclude that the chiral suppression structure in these models is
well-motivated by minimality.%
\footnote{In many models, these mixing masses are proportional to the quark masses
in order to preserve minimal flavor violation \cite{Chivukula:1987py},
which again leads to a chiral suppression.}

\subsection{Direct Detection of Dark Matter}
Direct detection experiments look for nuclear recoil events due to galactic dark matter.  
With standard assumptions about the dark matter halo distribution, the limits on signal
events are interpreted as limits on the DM-nucleus elastic recoil cross section.  
For a review of direct detection theory see \cite{Jungman}.  
A useful method to determine the nuclear recoil cross section is to find the effective
Lagrangian by integrating out the mediating particle, here $Q$ and computing the lowest 
order cross section.  
The connection between effective operators and the direct detection signal rate has been 
emphasized in recent work \cite{Fan:2010gt, Fitzpatrick:2012ix}.
Matrix elements of the effective operators are typically either proportional to the spin 
of the nucleon or add coherently for each nucleon, and for this reason
cross sections arising from  these operators are called spin dependent (SD) or spin independent (SI), respectively.  
The coherent scattering for spin-independent cross sections leads to an $A^2$ 
enhancement of $\mathcal{O}(10^4)$, producing the most stringent limits.  
In our results, we present results for SI interactions only, as we find that they 
provide stronger limits than SD interactions.   
Constraints from SI couplings are also less sensitive than those from SD couplings
to the flavors of the quarks that the dark matter interacts with.  
In all of our models, SI interactions are generated at some level, allowing us 
to focus on the current best SI limits from XENON100's 225 
$\text{kg}\cdot \text{day}$ run \cite{Aprile:2012nq}.  
To see how this will improve in the future, we also will add projected 
sensitivities for LUX and XENON1T taken from {\tt DMtools} 
\cite{DMtoolsProjections}.

Interestingly, current direct detection constraints are so strong that they put
some of the models well beyond the reach of the LHC.
For SI nucleon scattering, the quark operator with the largest matrix element  
is the vector current.  
By Lorentz invariance, the effective interaction coupling to the quark vector 
current can be a vector or pseudovector in the dark matter sector, with only the 
vector-vector coupling being unsuppressed in the nonrelativistic limit.  
Whether a dark matter vector operator is allowed depends on the dark matter 
quantum numbers.   
Complex dark matter models cannot forbid this vector coupling and thus have 
stringent constraints from XENON100 requiring multi-TeV masses.   
In models where the dark matter is its own antiparticle, the vector dark matter
operators vanish or are a total derivative.  Integrating by parts the total derivative changes the quark vector current into 
the mass operator which has a smaller matrix element.  
Thus in these models, much smaller masses for the dark matter are allowed.

\subsection{Heavy Partner Limit\label{sec:effint}}
If the quark partners are much heavier than the dark matter particle,
then both freeze-out and direct detection can be described by
contact interactions between dark matter and standard model
particles obtained by integrating out the quark partners.
Even for $m_\chi \sim m_Q$, annihilation is still described to
a reasonable approximation by this contact interaction because
the freeze-out temperature is $T \sim m_\chi / 25$.
It can break down for direct detection however because the 
quark partner can go on shell in the limit $m_Q \to m_\chi$.
Nonetheless, the structure of the effective interactions
determines many important features of freeze-out and direct detection scattering 
rates, so we will describe them here.
The main results are summarized in  Table~\ref{table:results}.
Our discussion in this section will be qualitative.
Precise formulas that are valid
for general parameters are given in Appendix B.

\paragraph{Fermion dark matter:}
In this case, $Q$ is a scalar.
This is similar to SUSY, where $Q$ is a squark.
Integrating out the scalar $Q$ gives the effective interaction
\beq\eql{fermionDMop}
\scr{L}_\text{eff}  
\sim \frac{\la^2}{m_Q^2} (\chi^\dagger \bar{\si}^\mu \chi) (q^\dagger \bar{\si}_\mu q),
\eeq
where we have used a Fierz identity.

If $\chi$ is a Majorana fermion, then $\chi^\dagger \bar{\si}^\mu \chi$
is pure axial vector, while the quark current is vector minus
axial vector.  
For the annihilation that sets the relic abundance, 
the $s$-wave cross section is 
suppressed by $m_q^2 / m_\chi^2$, so the annihilation is dominantly $p$-wave.  
This can be understood from $C$ and $P$ symmetries as reviewed in 
\cite{Bell:2008ey}.  
For any fermion-antifermion pair, $C=(-1)^{L+S}$, $P=(-1)^{L+1}$.  
For a Majorana pair, $C=+1$, so the $s$-wave channel requires zero net spin, 
$S=0$, which gives $P=-1$ and thus a pseudoscalar initial state $J^{PC}=0^{-+}$.  
In our models, the dark matter only couples to left-handed quarks 
(and right-handed anti-quarks), so in the limit of zero quark mass, the final spin 
is $S=1$ giving $C=(-1)^{L+1}$, $P=(-1)^{L+1}$ and thus $C$ and $P$ are
not conserved for any value of the final state $L$.     
Thus, $s$-wave annihilation into the quark-antiquark pair requires a helicity 
flip proportional to $m_q$.  
The $p$-wave does not suffer from such a suppression since the initial values 
are $S=1, L=1, C=+1, P=+1$.  
Thus, parametrically the annihilation cross section goes as
\beq
\text{Majorana DM:}\qquad
\si_\text{ann} \sim \la^4 \left(\frac{m_q^2}{m_Q^4} + v^2 \frac{m_\chi^2}{m_Q^4}\right).
\eeq

For direct detection, the non-relativistic regime determines the size of the scattering cross section.  
With the interaction \Eq{fermionDMop}, the 
coupling to the axial component of the quark current gives a spin-dependent
operator, while the mixed vector-axial
coupling is suppressed in the non-relativistic limit.  
Taking into account the next order correction in the momentum dependence of 
the $Q$ propagator leads to spin-independent operators, 
$m_q q q^c$ and a twist-two quark operator.  
These operators have smaller matrix elements than the quark vector current,
and therefore generally give weaker constraints.
An important exception discussed in \S\ref{sec:degenerate} is that
there is an enhancement in the degenerate limit due to the resonance when 
$m_\chi \sim m_Q$.   
Away from this limit, the direct detection constraints are easily
satisfied for masses that can be probed at the LHC.
Summarizing, 
for large $m_Q$, and assuming that the $s$-wave cross section is subdominant 
in the annihilation, the spin-independent cross section 
goes as
\beq
\text{Majorana DM:}\qquad 
\si_\text{SI} \sim \lambda^4 \frac{m_p^4 m_\chi^2}{m_Q^8}\sim \frac{m_p^4}{m_Q^4}\si_\text{ann},
\eeq
where $m_p$ is the proton mass.

If $\chi$ is a Dirac fermion, then the requirement of  
$C=+1$ no longer holds in the discussion of the annihilation
that produces the relic abundance.
This allows a vector-vector coupling which gives $s$-wave annihilation
\beq
\text{Dirac DM:}\qquad
\si_\text{ann} \sim \la^4 \frac{m_\chi^2}{m_Q^4},
\eeq
and an unsuppressed  spin-dependent interaction for direct detection of order
\beq
\text{Dirac DM:}\qquad
\si_\text{SI} \sim \la^4 \frac{m_p^2}{m_Q^4} \sim \frac{m_p^2}{m_\chi^2} \si_\text{ann}.
\eeq
Obtaining the correct relic abundance requires
$\si_\text{ann} \sim \text{pb}$, which given XENON100 limits  \cite{Aprile:2012nq} requires
heavy dark matter $m_\chi \gsim 5\TeV$, well out of the reach of LHC, or  very light dark matter, $m_\chi \lsim 10\GeV$.

\paragraph{Scalar Dark Matter:}
In this case $Q$ is a Dirac fermion.
Integrating out $Q$, we obtain an effective interaction
\beq
\scr{L}_\text{eff} \sim \frac{\la^2}{m_Q^2} \chi^\dagger 
q^\dagger i \bar{\si}^\mu \d_\mu(\chi q)
= \frac{\la^2}{m_Q^2} \chi^\dagger \d_\mu \chi
q^\dagger i \bar{\si}^\mu q + O(m_q).
\label{eq:scalareffint}
\eeq

If $\chi$ is a real scalar, we can integrate by parts to write the
interaction as
\beq
\text{Real scalar DM:}\qquad
\scr{L}_\text{eff} \sim \frac{\la^2}{m_Q^2} \chi^2 \d_\mu 
(q^\dagger \bar{\si}^\mu q)
\eeq
The divergence of the left-handed quark current is proportional
to $m_q$, which gives a suppression for light quarks.
This means that the annihilation is always chirally suppressed and thus $s$ and $p$-wave are both chirally suppressed.  
The coupling to the quark mass operator also means that direct detection is suppressed.  This gives the parametric scaling 
\beq
\text{Real scalar DM:}\qquad
\si_\text{SI} \sim \lambda^4 \frac{m_p^4}{m_\chi^2 m_Q^4} \sim \frac{m_p^4}{m_q^2 m_\chi^2} \si_\text{ann}.
\eeq
This leads to strong constraints unless the top quark is kinematically 
accessible in dark matter annihilation.  

If $\chi$ is a complex scalar, one can integrate by parts to find
\beq
\text{Complex scalar DM:}\qquad
\scr{L}_\text{eff} \sim \frac{\la^2}{2 m_Q^2} \bigl(\chi^\dag \!\!\stackrel{\leftrightarrow}{\d}_\mu \! \chi \bigr)
(q^\dagger \bar{\si}^\mu q) + O(m_q).
\label{eq:complexscalareffint}
\eeq
For annihilation, the dark matter 
operator vanishes in the limit $v\to 0$, so the annihilation remains chirally suppressed at $s$-wave, but not in the $p$-wave. 
The direct detection is unsuppressed since the interaction in 
Eq.~(\ref{eq:complexscalareffint}) has a nonzero vector-vector component.  
This leads to the scaling
\beq
\text{Complex scalar DM:}\qquad
\si_\text{SI} \sim \lambda^4 \frac{m_p^2}{m_Q^4} \sim \frac{m_p^2}{m_\chi^2}\si_\text{ann},
\eeq
which again requires dark matter mass in the multi-TeV range or very light GeV range.

\paragraph{Vector Dark Matter:}
In this case $Q$ is again a Dirac fermion.
Integrating out $Q$, we obtain an effective interaction
\beq
\scr{L}_\text{eff} \sim \frac{\la^2}{m_Q^2}
q^\dagger \bar{\si}^\mu \chi_\mu^\dagger i \si^\nu \d_\nu
\left( \chi_\rho \bar{\si}^\rho q \right).
\eeq
We can use the identity
$\bar{\si^\mu} \si^\nu \bar{\si}^\rho 
= g^{\mu\nu} \bar{\si}^\rho-g^{\mu\rho} \bar{\si}^\nu+g^{\nu\rho} \bar{\si}^\mu 
-i \epsilon^{\mu\nu\rho\kappa}\bar{\si}_\kappa$ \cite{Dreiner:2008tw} 
to simplify this.
For a real vector
$\chi_\mu^\dagger = \chi_\mu$, so we obtain 
\beq
\text{Real vector DM:}\qquad
\scr{L}_\text{eff} \sim \frac{\la^2}{m_Q^2} \left[
i \chi^\mu \chi^\nu q^\dagger \bar{\si}_\mu 
\!\!\stackrel{\leftrightarrow}{\d}_\nu\! q 
+ 
\tilde{F}_{\mu\nu} \chi^\mu q^\dagger \bar{\si}^\nu q \right] + O(m_q).
\eeq
For the relic abundance, the first term contains a twist-two quark interaction which has an unsuppressed $s$-wave contribution.  
For direct detection, taking the non-relativistic limit, this twist-two component has a small matrix element, but again is enhanced near degeneracy.  The second term does not give a large vector-vector interaction since the $\nu=0$ term is proportional to $\tilde{F}_{\mu0} = \epsilon_{\mu 0 \rho \si} \d^\rho A^\si,$ which is suppressed by the momentum transfer of the dark matter.  In addition, that term gives a spin-dependent interaction when $\nu$ is a spatial index.  Given the unsuppressed relic abundance and the suppressed direct detection, we find 
\beq
\text{Real vector DM:}\qquad
 \si_\text{SI} \sim  \lambda^4 \frac{m_p^4}{m_\chi^2 m_Q^4} \sim \frac{m_p^4}{m_\chi^4} \si_\text{ann}.  
\eeq

For a complex vector, using the sigma matrix identity one finds an allowed vector-vector coupling, $A^\dag_\mu \d_\nu A_\mu \, q^\dag \bar{\si}^\nu q$, which gives unsuppressed rates for both annihilation and direct detection.  This gives  
\beq
\text{Complex vector DM:}\qquad \si_\text{SI} \sim  \lambda^4 \frac{m_p^2}{m_Q^4} \sim  \frac{m_p^2}{m_\chi^2}\si_\text{ann}
\eeq
and again pushes the dark matter mass to several TeV or below 10 GeV.

\begin{table}
\begin{center}
\begin{tabular}{|c|c|c|c|}
\hline
  \multicolumn{2}{|c|}{Model} & \multirow{2}{*}{{Relic Abundance}} &  \multirow{2}{*}{\parbox{3cm}{Direct Detection}} \\
\cline{1-2}
  $\chi$ & Q & &     \\ 
\hline
 Majorana fermion& Complex scalar& \begin{tabular}[x]{@{}c@{}} $a \sim m_q^2$\\$ \la \sim 0.5-2$\end{tabular}&  \begin{tabular}[x]{@{}c@{}}  Suppressed  \\ $\si_\text{SI}\overset{m_Q\gg m_\chi}{\sim}  \frac{m_p^4}{m_Q^4}\si_\text{ann} $\end{tabular}    \\
\hline
Dirac fermion & Complex scalar &  $\la \sim 0.2-1$ &  \begin{tabular}[x]{@{}c@{}} Unsuppressed\\ $\si_\text{SI} \overset{m_Q\gg m_\chi}{\sim}  \frac{m_p^2}{m_\chi^2}\si_\text{ann} $\end{tabular}\\
\hline
Real scalar & Dirac fermion& \begin{tabular}[x]{@{}c@{}} $a,b \sim  m_q^2 $\\$ \la \sim 0.5-5$\end{tabular} &  \begin{tabular}[x]{@{}c@{}} Suppressed if $m_\chi > m_t$ \\ $\si_\text{SI} \overset{m_Q\gg m_\chi}{\sim} \frac{m_p^4}{m_q^2 m_\chi^2}\si_\text{ann} $\end{tabular}\\
\hline
Complex scalar & Dirac fermion &  \begin{tabular}[x]{@{}c@{}} $a \sim m_q^2 $\\$ \la \sim 0.5-2$\end{tabular} &   \begin{tabular}[x]{@{}c@{}}Unsuppressed\\  $\si_\text{SI} \overset{m_Q\gg m_\chi}{\sim}  \frac{m_p^2}{m_\chi^2}\si_\text{ann} $\end{tabular}\\
\hline
Real vector & Dirac fermion &$\la \sim 0.05-0.5$ &  \begin{tabular}[x]{@{}c@{}}Suppressed\\ $\si_\text{SI}\overset{m_Q\gg m_\chi}{\sim}   \frac{m_p^4}{m_\chi^4}\si_\text{ann} $\end{tabular} \\
\hline
Complex vector & Dirac fermion &$\la \sim 0.07-0.7$ &  \begin{tabular}[x]{@{}c@{}}Unsuppressed\\ $\si_\text{SI}\overset{m_Q\gg m_\chi}{\sim}  \frac{m_p^2}{m_\chi^2}\si_\text{ann} $\end{tabular} \\
\hline
\end{tabular}
\begin{minipage}{5.5in}
\caption{Overview of results for relic abundance and direct
detection for the various models. \label{table:results}}
\end{minipage}
\end{center}
\end{table}

\subsection{Indirect Detection}
Indirect detection experiments looking for dark matter annihilation or decay products in cosmic rays are another potential constraint on these models.  
Our models do not have Sommerfeld enhancement, 
so the annihilation cross section today is smaller than the thermal annihilation cross section $\langle \si_\text{ann} v\rangle = 3 \times 10^{-26}\; \text{cm}^3/\text{s}$ required for relic abundance.
For the most part, indirect detection constraints on dark matter annihilation 
channels  give upper bounds for cross sections that are larger than the
thermal value, so these do not constrain our models.
One exception is gamma ray limits from Fermi-LAT observation of dwarf galaxies.  
By stacking several observed galaxies, a limit stronger than the thermal cross section 
can be achieved for certain annihilation channels.  
The one that applies to our models is the constraints on the cross section 
for annihilation to $b\bar{b}$ pairs, which has been analyzed in 
\cite{GeringerSameth:2011iw, Ackermann:2011wa}.
The precise limit on the dark matter mass for a thermal cross section with
100\% annihilation to $b\bar{b}$ pairs has large uncertainties due to the 
dark matter distribution of these galaxies, but gives lower limit of about 60~GeV.  

As we will see below, this
region is already highly constrained by direct detection and collider
searches in our models.
In addition, for models where the dark matter couples to all quarks, the presence
of other annihilation channels significantly weakens this bound.
If the annihilation cross section $s$-wave component is not chirally suppressed 
(as in the Dirac fermion and real vector dark matter models), then the 
annihilation currently is evenly spread amongst all of the open channels, 
so that the $b\bar{b}$ rate is $\frac 15$--$\frac 16 \times$ thermal.  
Whereas if the $s$-wave component is chirally suppressed and the $p$-wave is not 
(as in the Majorana fermion and complex scalar models), 
the $s$-wave is dominantly into the heaviest quarks available, 
but this is typically smaller than thermal since the $p$-wave cross section at 
freezeout was important in getting a large enough cross section for the relic abundance.  Finally, for the real scalar model, both $s$ and $p$-waves are chirally suppressed.  
In this case, if the dark matter mass is between the bottom and top quark masses, 
it will have a nearly thermal cross section to annihilate into bottom quarks.  
Then we would have a dark matter lower limit from the stacked analysis, requiring 
$m_\chi \gsim 60$ GeV.  
Finally, if the dark matter only couples to third generation quarks, we also expect to find a similar limit of $m_\chi \gtrsim 60$ GeV.

\subsection{Limits from Collider Experiments}
Since the new particles are odd under a $Z_2$ parity, 
they are produced in pairs in colliders.  
Thus the primary production channels at the LHC are $pp \rightarrow Q Q^\dagger$, $QQ$, 
$Q \chi$, $\chi \chi^\dagger$.  
Since $Q$ decays to $q\chi$, these channels produce signatures with 
missing transverse energy (MET) and 2,1, and 0 parton-level jets, respectively.  
The zero jet event would be invisible to the detector, but an additional initial 
state radiation jet can make $p p \rightarrow \chi \chi + $ jet a visible signal.  

This leads to two primary detection signals: 
$\text{dijet} + \text{MET}$ and monojet + MET.  
For the $\text{dijet} + \text{MET}$ signal, we utilize two CMS simplified model searches for light generation squarks  
based on 11.7 fb$^{-1}$ \cite{CMSbottom} and 19.5 fb$^{-1}$ \cite{CMS-PAS-SUS-13-012} luminosity at 8 TeV which have cross section limits as a function of squark, neutralino mass.  These are combined by taking the best limit of the two searches and will be described from now on as the CMS dijets+MET search.  We will also use the bottom squark limits from the earlier analysis \cite{CMSbottom}.  We also found that the latest CMS search for top squarks  \cite{CMS-PAS-SUS-13-011} sets similar limits to the bottom squark search, so we chose to omit it from our plots.  For monojet sensitivities, we used the latest CMS search \cite{CMS-PAS-EXO-12-048}, which placed limits on the monojet cross section in different MET bins.  For the monojet and light generation squark search, we only consider light quark and gluon final states as jets.
 
To determine the cross sections for our models, the event rates for the collider production of the quark partners was calculated at parton level at leading order using {\tt MadGraph5} v1.4.8.4 \cite{Alwall:2011uj}.%
\footnote{The production cross sections obtained from {\tt MadGraph} can vary up to 
about $25$--$50\%$ depending on the factorization/renormalization scale used. 
We do not know the best scale given the new contributions due to the 
$\la$ interaction, so we have used the default scale in {\tt MadGraph}.} 
The {\tt MadGraph} model files were generated using 
{\tt FeynRules} v1.6.0 \cite{Christensen:2008py}.  
Since our trilinear interaction strength can be substantial, this allows us to take into account the important production mechanism of same sign quark partners via a $t$-channel exchange of the dark matter particle as well as the monojets due to associated production of $Q \, \chi$.  To apply the CMS limit for squarks to our total cross section, we assume that the signal efficiencies for QCD production and the new $t$-channel process are similar.  Close to the degeneracy line, initial state radiation plays a crucial role for the signal selection.  Initial state radiation differs between gluon and quark initial states and thus the naive approach of applying the simplified model cross section limit could break down in this part of the parameter space.      

The diagrams that contribute to the monojet signal are shown in Fig.~\ref{fig:monojet}.  The $Q$ particle is pair produced at colliders primarily through gluon-gluon and quark-antiquark annihilation (see Fig.~$\ref{fig:jetsMET}$). Production through strong production is generally larger since the gluon dominates in the proton PDF.  The exception is at large quark partner masses, where the up and down quark PDFs become larger than the gluon.  For the models where same sign quark partners can be produced by a $t$-channel exchange of the dark matter (similar to the third diagram of Fig.~$\ref{fig:jetsMET}$ with $q q \to Q Q$ production instead), this enhances the production rate at large quark partner masses.  Thus, the new channel allow our limits to extend beyond the CMS analysis, even though we are just using a leading order calculation.

\subsection{Near Degenerate Effects\label{sec:degenerate}}

Near the degeneracy of $m_Q$ and $m_\chi$ direct detection rates can gain additional sensitivity.  The enhancement of the direct detection cross section can be seen by considering tree-level interactions with the nucleus.  As shown in Fig.~\ref{fig:basicint}, nucleon scattering in the $s$-channel process leads to resonant enhancement when $\sqrt{s} \simeq m_\chi$ is close to $m_Q$.   As emphasized in \cite{Hisano-wino}, this enhances the coefficients of the spin-independent operators,   increasing their direct detection rates.    Conversely, the collider searches lose sensitivity near degeneracy.  In this regime, the small mass splitting leads to soft jet production and signal events are lost due to $p_T$ cuts.  Hence, we expect direct detection to be complementary to collider bounds in this region.  

Another signal that is enhanced in this region is the indirect detection signal of photons produced in dark matter collisions with protons in AGN jets \cite{Gorchtein:2010xa, Chang:2012sk}.  However, \cite{Chang:2012sk} showed that even when saturating the XENON100 bound and taking favorable AGN and dark matter parameters, that this signal was still out of current sensitivity of gamma ray telescopes.    

Finally, for the relic abundance calculation, 
when $m_Q \simeq m_\chi$, both particles freeze out at approximately 
the same temperature, leading to coannihilation effects not considered in 
our analysis.  
Since the squarks tend to have stronger annihilation cross sections,
this reduces the required $\la$ near degeneracy \cite{Griest:1990kh},
weakening the direct detection limits.
Rather than perform a detailed analysis of this special region, we 
will simply highlight this region where our approximations begin to break down
and omit the region $m_Q < 1.1\, m_\chi$ from our results.

\section{Results \label{sec:Results}}
We now present our results.  
The main conclusions of \S\ref{sec:Constraints}
are summarized in Table~\ref{table:results}.  
As described earlier, due to the constraint 
$\si_\text{SI} \lesssim \text{few}\times 10^{-45}\, \text{cm}^2$
from XENON100 \cite{Aprile:2012nq} 
the models where spin-independent interactions are not suppressed 
have  $\si_\text{SI} \sim (m_p/m_\chi)^2 10^{-36}\, \text{cm}^2$,
which requires dark matter masses above a TeV or below 10 GeV.  
The light mass region is constrained by the CMS monojet search, leaving the multi-TeV range masses as the only unconstrained region.  Thus, over the parameter space, the Dirac fermion, complex scalar, and complex vector DM models are primarily probed by direct detection experiments and are essentially irrelevant for the LHC.  
We therefore focus on the remaining models, which can be probed
both by the LHC and direct
detection experiments.

\subsection{Majorana dark matter}
\begin{figure}[h!] 
\centering \includegraphics[scale=0.45]{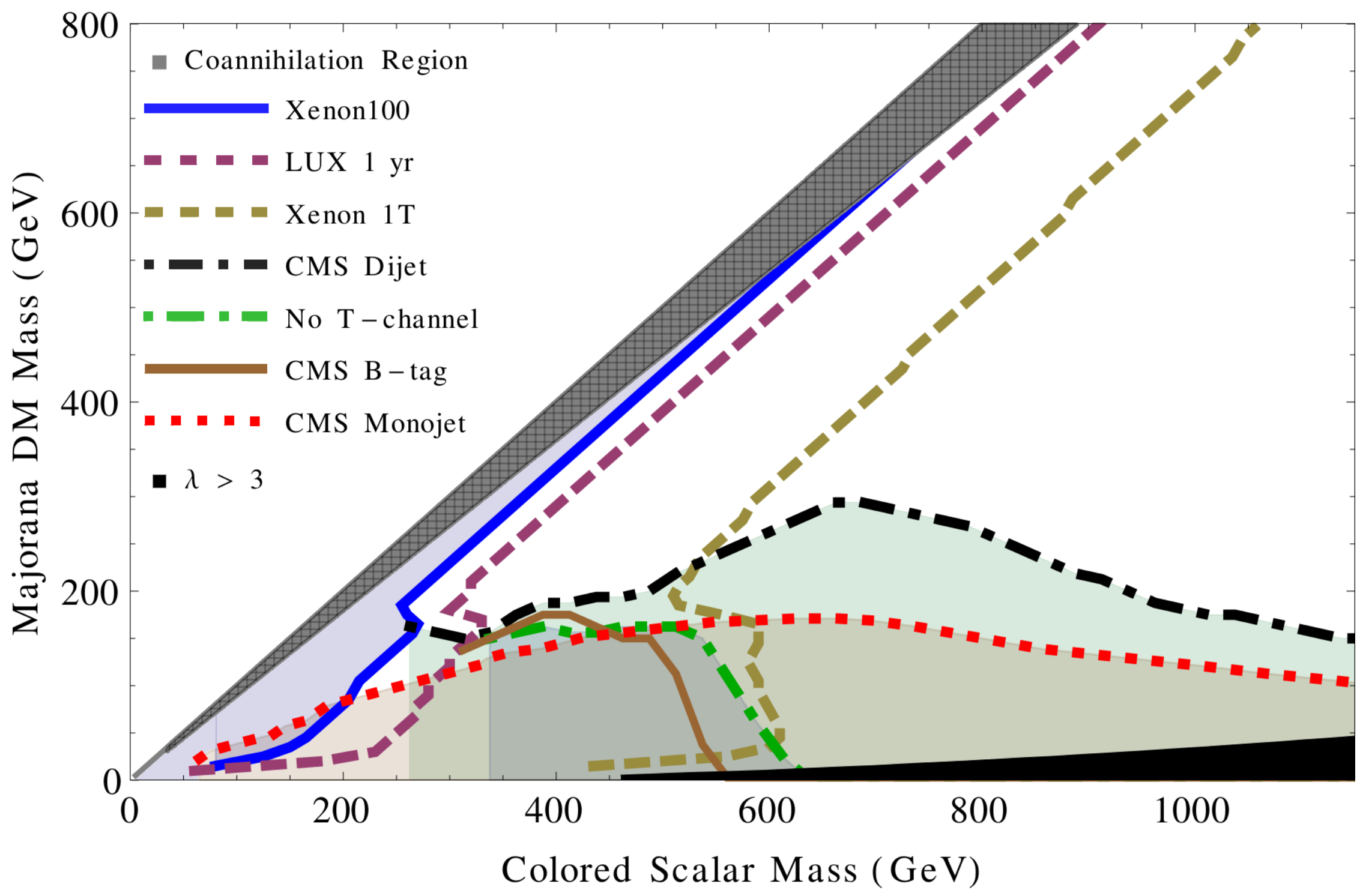}
\begin{minipage}{5.5in}
\caption{\small 
Limits on Majorana dark matter coupling to all generations.  
The limits from the CMS dijet searches are shown with lines (black dot dashed, green dot dashed, brown solid) taking into account the production
modes (all, QCD only, bottom quark) and the CMS monojet is shown in red dotted.
The direct detection limits (XENON100 in blue solid, projected LUX and XENON1T in dashed) have an edge at $m_\chi \simeq m_t$ due to the 
effects of the top quark on the relic abundance.
There are two regions where the results have large uncertainties.  
In the grey region $m_Q < 1.1\, m_\chi$,  coannihilation effects can strongly suppress
$\la$, weakening the bounds.
In the black region $m_Q \gg m_\chi$, $\la > 3$ is required to obtain the 
correct relic abundance.}
\label{M1F1}
\end{minipage}
\end{figure}

\begin{figure}[h!] 
\centering \includegraphics[scale=0.45]{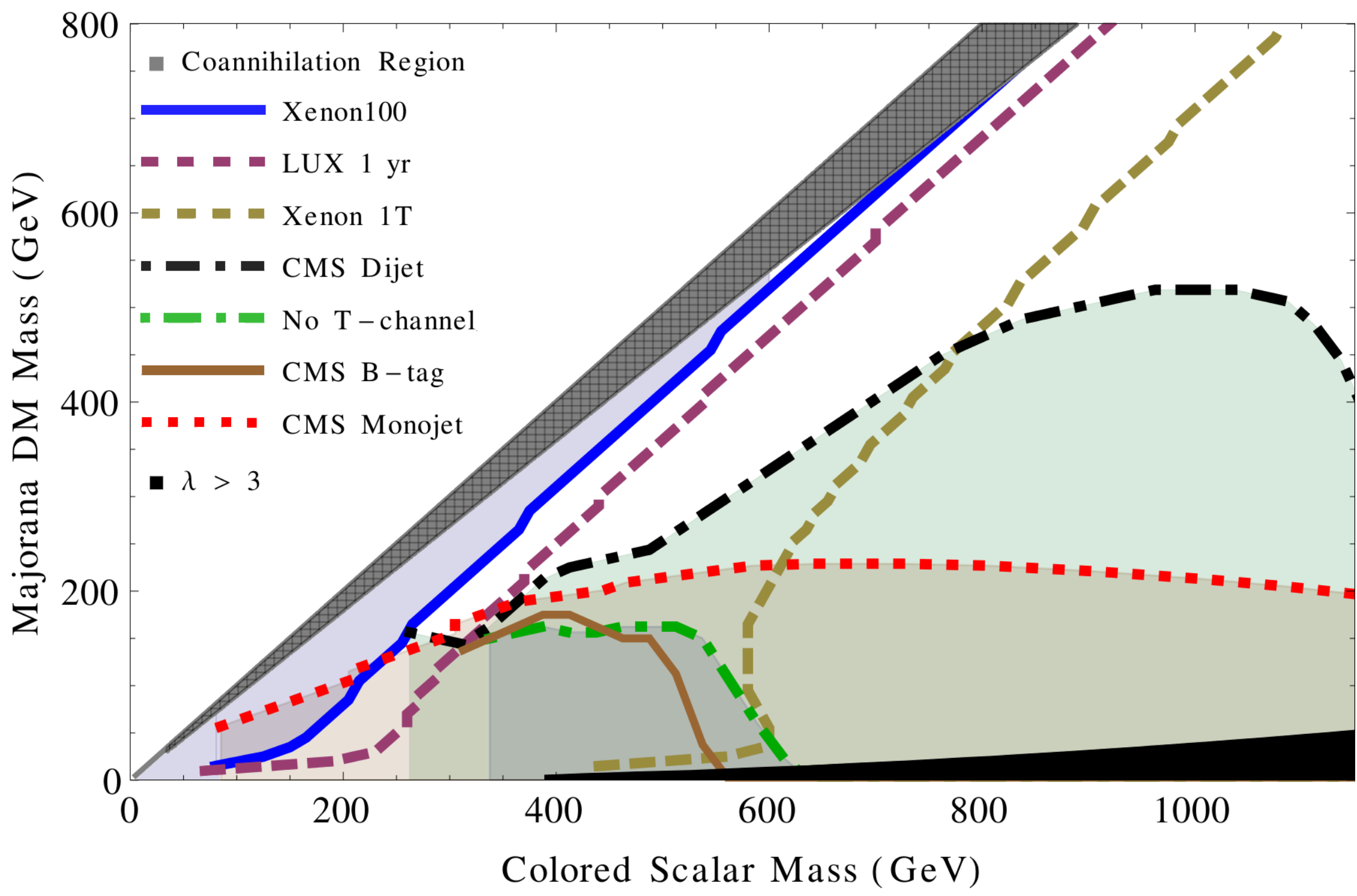}
\begin{minipage}{5.5in}
\caption{\small 
Limits on Majorana dark matter coupling to the lightest two generations.  Labeling as in Fig.~\ref{M1F1}.}
\label{M1F4}
\end{minipage}
\end{figure}

We begin with the case where the dark matter is a
Majorana fermion.
The results for the models where the dark matter couples to all generations and just the light quarks
are presented in Figs.~\ref{M1F1} and \ref{M1F4}.
We see that these models have a large region of parameter space allowed
by current constraints.

There are several important features to note in these results.
First, the CMS $\text{dijet} + \text{MET}$ search gives the most stringent
constraint, although the monojet searches are also sensitive.
Note that the sensitivity extends to partner masses well above 1 TeV.
Our results are cut off at $1.2$~TeV because the CMS dijet search does
not present results for quark partner masses above this value, 
even though they clearly have sensitivity there.
The reason for the sensitivity to very large masses is that for large 
$m_Q$ the coupling $\la$ is getting large in order to produce the correct
relic abundance.
The coupling is still perturbative in most of the interesting region, however.
In the plots this is indicated by a black region where $\la > 3$, corresponding
to a perturbative expansion parameter $\la^2 / 8\pi^2 \sim 0.1$.
The bound from the $\text{dijet} + \text{MET}$ is strengthened considerably
by the presence of the $t$-channel contribution proportional to $\la^2$.
This is because Majorana dark matter allows $t$-channel 
production from the $qq$
(as opposed to $q\bar{q}$) initial state, which has a large PDF at
large $x$.
The dijet bounds we present are obtained by applying the cross section limit
for the simplified model considered in the CMS analysis.  We expect the sensitivity differences between QCD production and the new $t$-channel process to be most pronounced when the quark partner and dark matter are nearly degenerate due to the differences in initial state radiation off of gluons and quarks.
Thus, we advocate that CMS and ATLAS present results for a simplified model
with $t$-channel production to get accurate limits on this scenario.

The monojet searches are less sensitive, but they
have substantial overlap with the $\text{dijet} + \text{MET}$ search.
The monojet limits are slightly weaker than the dijet limits but also extend to 
large values of $m_Q$, where the limits asymptote to the `effective' dark matter' 
approach.
In our models, the monojet signal will have a broad
peak in the missing transverse energy spectrum around $\frac{m_Q^2-m_\chi^2}{2m_Q}$,
compared with the effective dark matter models, which have a falling, featureless enhancement.
It may be interesting to investigate whether searches optimized for
the models considered here will have significantly enhanced sensitivity.

\begin{figure}[h!] 
\centering \includegraphics[scale=0.45]{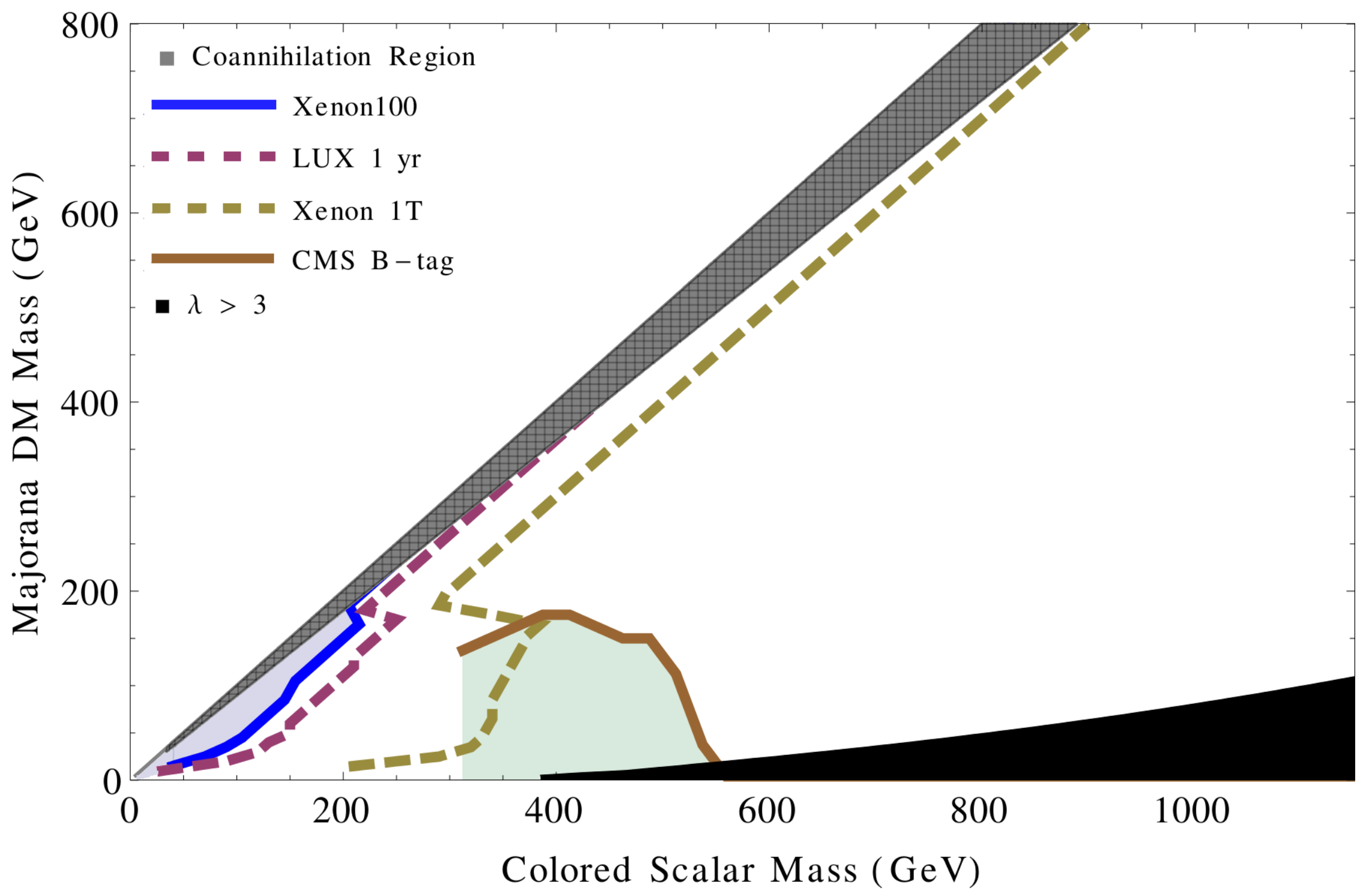}
\begin{minipage}{5.5in}
\caption{\small 
Limits on Majorana dark matter coupling to third generation only.  Labeling as in Fig.~\ref{M1F1}.}
\label{M1F2}
\end{minipage}
\end{figure}

For the case where dark matter couples only to third-generation quarks,
production is due to QCD only and thus
is more suppressed.
The bottom quark partner results are shown in Fig.~\ref{M1F2}.
(We also show the results of the third generation search for the case where
the dark matter couples to all 3 generations in Fig.~\ref{M1F1})
There is presently no `mono-$b$' search, but this would presumably be
quite sensitive in this model.

The collider limits become much weaker as we approach the degenerate
limit $m_\chi \simeq m_Q$, since this reduces the missing energy.
However, in the region direct detection has enhanced sensitivity,
because the energy denominator suppressing the direct detection cross
section is $m_Q - m_\chi$.
In fact, the current XENON100 limit already rules out the entire region near the 
degenerate limit.  
In this region, co-annihilation becomes important, and this was not included in 
our relic abundance calculations, so these limits are not fully reliable.
However, as can be seen in the XENON1T and LUX projections, the improvements in future 
years in spin-independent direct detection limits will push the sensitivity into a
region where the coannihilation effects are negligible.
The collider limits on the degenerate region are also expected to improve,
so in future years we may expect direct detection and collider searches to 
fully probe this region.

Note that the direct detection bounds are very weak for
$m_\chi \ll m_Q$.
This is due to the fact that the spin-dependent cross section
goes as $m_Q^{-4}$, as shown in Table~\ref{table:results}.
This feature is not present in the other models considered below,
so in these cases direct detection is more sensitive for $m_\chi \ll m_Q$.

\subsection{Real scalar dark matter}
\begin{figure}[t!] 
\centering \includegraphics[scale=0.45]{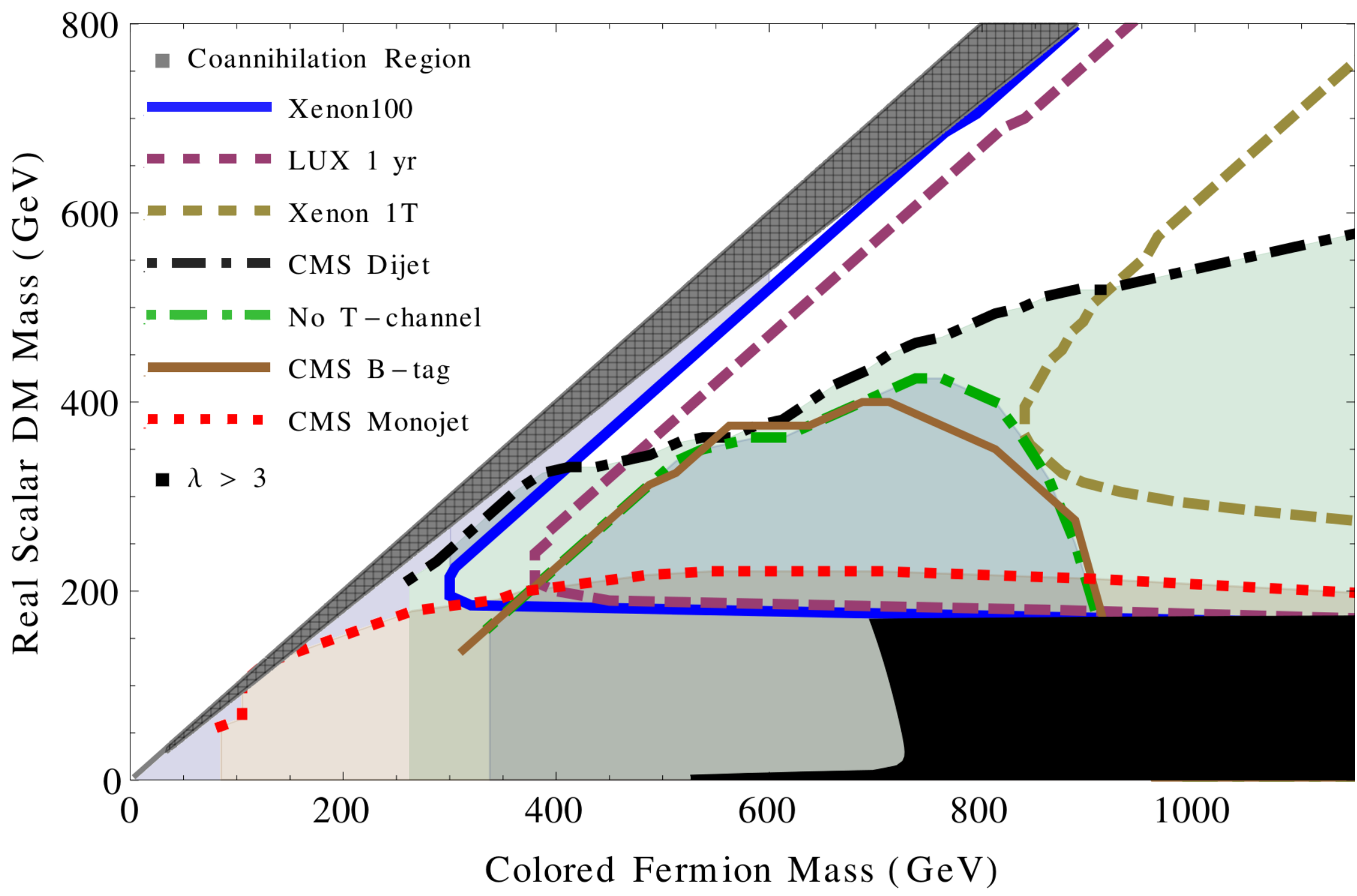}
\begin{minipage}{5.5in}
\caption{\small 
Limits on real scalar dark matter coupling to all generations.  Labeling as in Fig.~\ref{M1F1}.}
\label{M3F1}
\end{minipage}
\end{figure}

\begin{figure}[t!] 
\centering \includegraphics[scale=0.45]{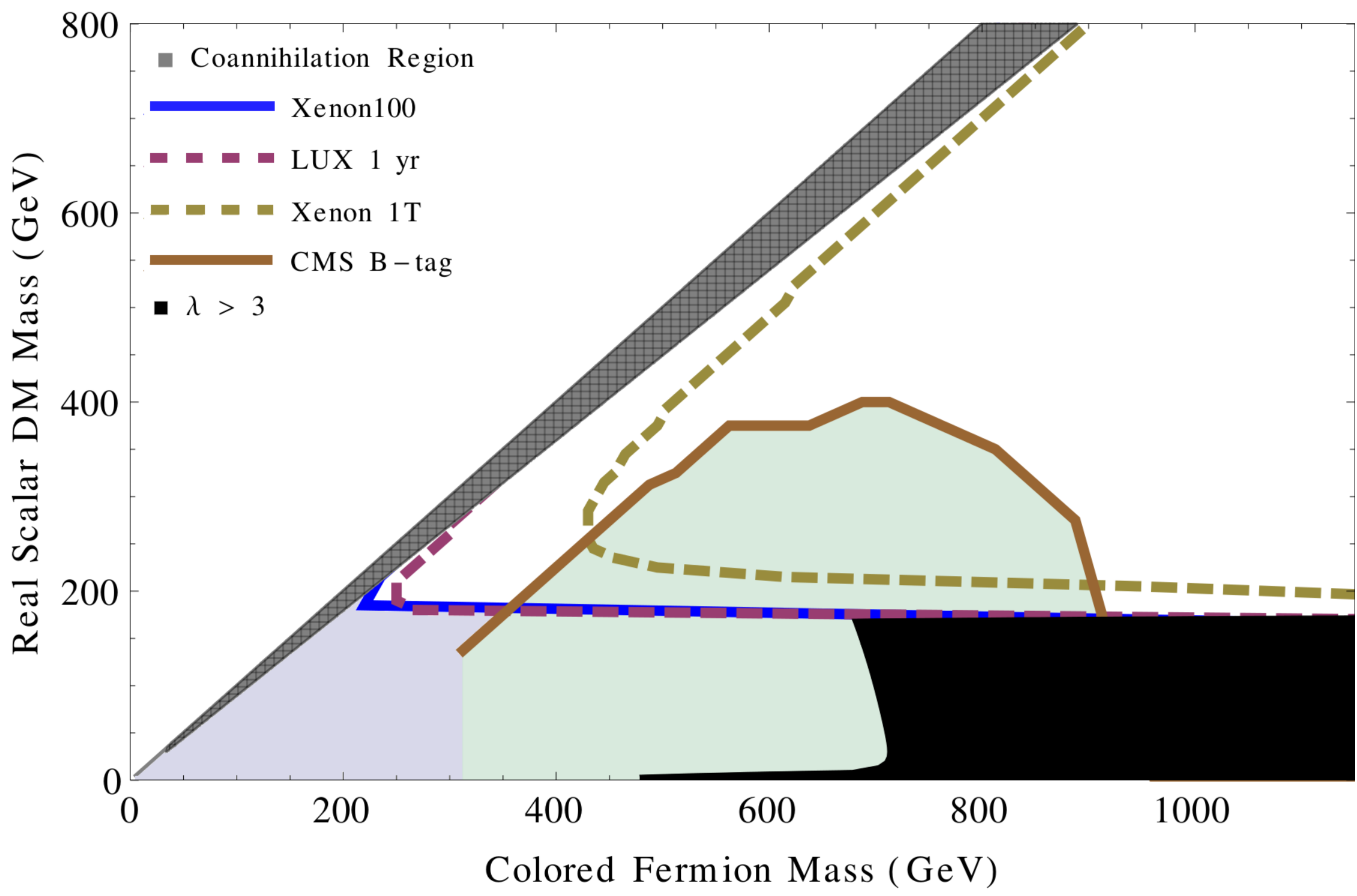}
\begin{minipage}{5.5in}
\caption{\small 
Limits on real scalar dark matter coupling to third generation only.  Labeling as in Fig.~\ref{M1F1}.}
\label{M3F2}
\end{minipage}
\end{figure}

For this model, both the $s$- and $p$-wave 
annihilation cross sections are chirally suppressed.
Therefore, if the dark matter couples only to the 
lightest two generations, its interaction strength is required to be
non-perturbatively strong to get the right relic abundance
unless $m_Q \lesssim 400$ GeV.  
However, this region is excluded by the XENON100 and CMS monojet limits.  
Thus, we present results only for the cases of coupling to all generations 
and the third generation only.     
The results are shown in Figs.~\ref{M3F1} and \ref{M3F2}.
If $m_\chi < m_t$, the coupling $\la$ cannot account for the relic
abundance unless $m_Q \lsim 700$~GeV.

Note that the CMS dijet limits are enhanced with respect to the Majorana models
because fermion quark partners have a larger production cross section than scalar 
quark partners.  
The constraints using just the QCD production mechanism would already rule out 
quark partners up to about 1 TeV for light dark matter.  
Including the $t$-channel, again extends the limit to higher masses.      

In the models where dark matter couples to all generations, the XENON100 
limit is comparable 
to the CMS monojet limit.  
This is a result of the relic abundance constraint:
the value of $\la$ required to get the right relic
abundance drops sharply once $m_\chi > m_t$.

\subsection{Real vector dark matter:}
\begin{figure}[h!] 
\centering \includegraphics[scale=0.45]{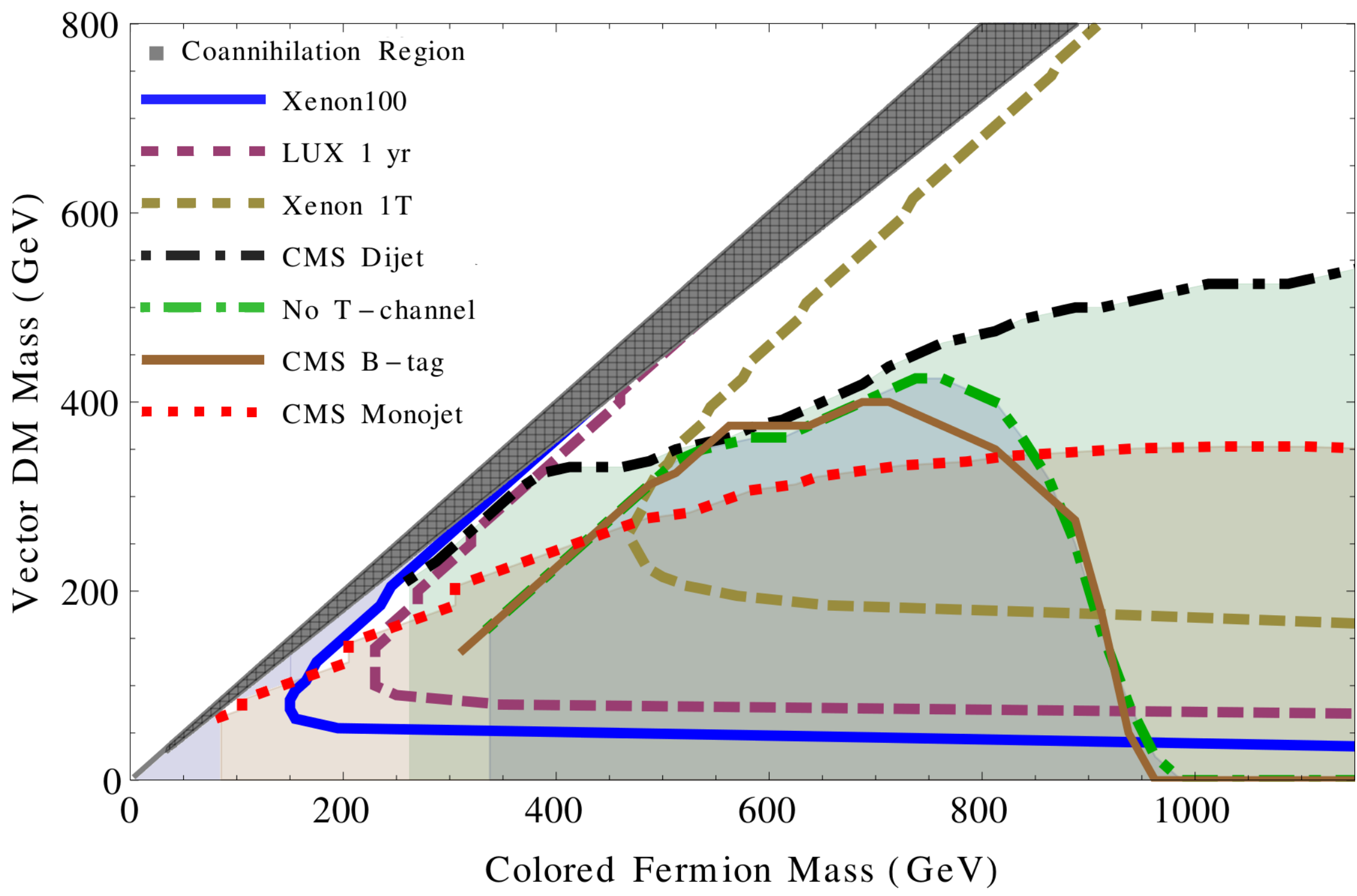}
\begin{minipage}{5.5in}
\caption{\small 
Limits on real vector dark matter coupling to all generations.  Labeling as in Fig.~\ref{M1F1}.}
\label{M5F1}
\end{minipage}
\end{figure}

\begin{figure}[h!] 	
\centering \includegraphics[scale=0.45]{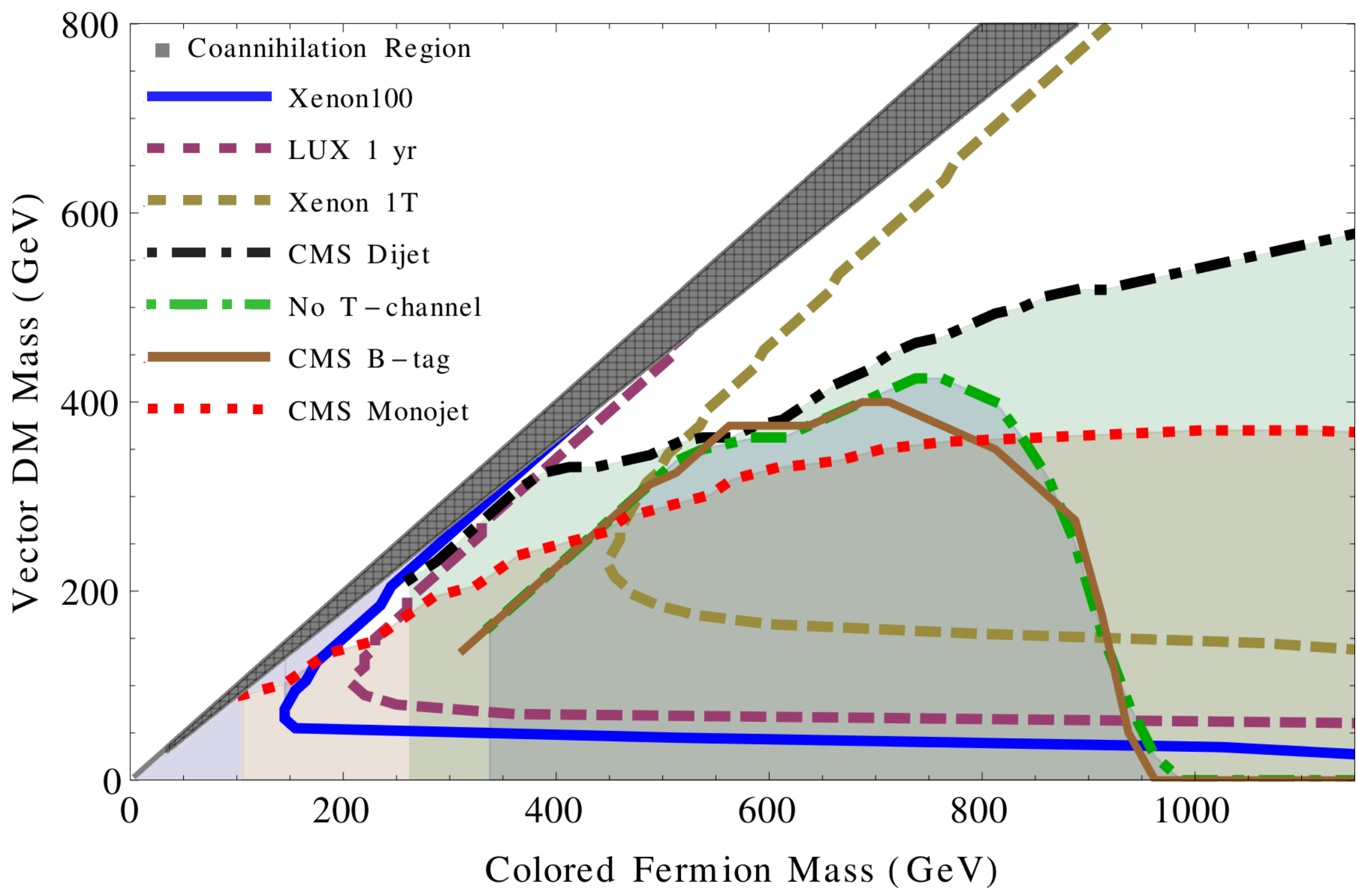}
\begin{minipage}{5.5in}
\caption{\small 
Limits on real vector dark matter coupling to the lightest two generations.  Labeling as in Fig.~\ref{M1F1}.}
\label{M5F4}
\end{minipage}
\end{figure}

\begin{figure}[h!] 
\centering \includegraphics[scale=0.45]{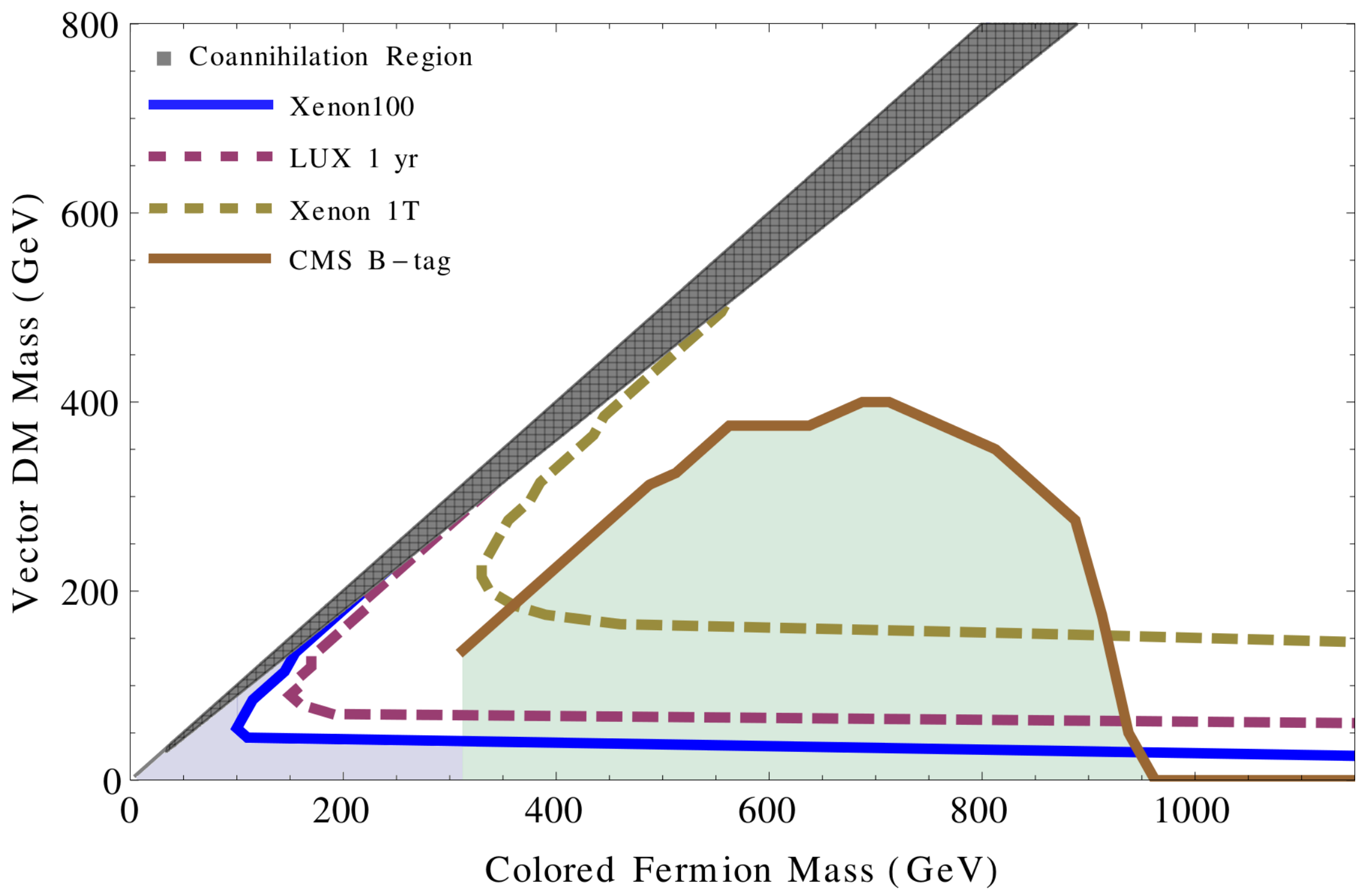}
\begin{minipage}{5.5in}
\caption{\small 
Limits on real vector dark matter coupling to third generation only.  Labeling as in Fig.~\ref{M1F1}.}
\label{M5F2}
\end{minipage}
\end{figure}

For the real vector dark matter model, the interaction strength is small, 
since neither the $s$ and $p$-wave cross sections are chirally suppressed.  
The results are in Figs.~\ref{M5F1}, \ref{M5F4}, and \ref{M5F2}.  
These smaller couplings lead to weaker direct detection constraints than the real 
scalar dark matter case.  
Note the behavior of an asymptotic limit as $m_Q \gg m_\chi$ is explained by the
fact that in this limit, $\si_\text{SI}$ only depends on $m_\chi$ 
(see Table~\ref{table:results}). 
On the other hand, the collider constraints are still strong due to the large cross 
section for fermion quark partners and the $t$-channel mechanism.  
The $t$-channel matrix element receives an enhancement of $\sim m_{Q}^2/m_{\chi}^2$ due to the $q^\mu q^\nu/m_\chi^2$ part of the dark matter propagator.  This enhancement will be cut off by the 
Higgs sector responsible for giving a mass to the dark matter vector particle,
and so the $t$-channel bound given here is too strong.
In a complete model, the collider limit will be somewhere between the
bounds with and without the $t$-channel contribution.
The monojet bounds are not affected by this theoretical uncertainty, 
and these extend to large values of $m_Q$.


\section{Conclusions \label{sec:Conclusions}}
We have proposed and studied a new phenomenological approach to
interpreting dark matter searches, 
based on a minimal particle content required to explain WIMP dark matter.
The models consist of a singlet dark matter particle coupling to quarks
and `quark partners.'
We consider dark matter with spin $0$, $\frac 12$, and $1$
that is or is not its own antiparticle.
In each case, 
imposing the constraint that the dark matter have the correct
thermal relic abundance fixes the interaction strength, so the model
is completely specified by the masses of the dark matter and quark partners.

These `effective WIMP' models therefore have a 
2-dimensional parameter space in which the reach of dark matter direct and indirect
detection experiments can be directly compared to collider searches for missing
energy events.
This extends the approach of `effective dark matter' models,
and is complementary to more ambitious approaches based on complete
models, such as scans for supersymmetric dark matter 
({\it e.g.} \cite{Cahill-Rowley:2013dpa}).

Our main results are as follows.
\begin{itemize}
\item
The most sensitive direct detection constraints come from spin-independent
interactions.
Indirect detection is currently not competitive.
\item
The most sensitive collider constraints are from
$\text{jets} + \text{MET}$ searches and monojet searches,
with the former generally more sensitive.
The production cross sections at colliders are greatly enhanced by processes
involving the new coupling, extending the reach for the colored
states to very high masses.
Collider searches for effective WIMPs may be improved by optimizing
for these production modes.
\item
The direct detection and collider constraints are remarkably complementary.
If the dark matter is not its own antiparticle, direct detection
constraints require the dark matter mass to be in the multi-TeV
range, far out of reach of LHC searches.
If the dark matter is its own antiparticle, both collider and direct
detection are sensitive.
Direct detection has enhanced sensitivity in the degenerate
region $m_Q \simeq m_\chi$ where collider searches are less sensitive
due to reduced missing energy.
On the other hand, collider searches generally probe a larger
region of the parameter space away from the degenerate limit.
\item
Mono-$b$ searches can significantly enhance sensitivity to models
where dark matter couples dominantly to third-generation quarks.
\end{itemize}

We advocate that these models can play an important role in
interpreting searches for astrophysical dark matter and 
dark matter at colliders.
They allow us to unambiguously compare both kinds of dark matter
searches in the context of well-defined physical minimality assumptions.
Of course these assumptions are strong ones, but they are crucial ones to test.
If a signal is observed in either direct detection or collider searches,
one of the most important questions to answer is whether the signal can be explained
by a minimal number of additional states, or whether the dark matter
is part of a larger sector of new particles (as in SUSY) that can be 
searched for at colliders.
We have shown that effective WIMP models have multiple overlapping 
as well as complementary probes that can be unambiguously compared
to help answer this question.

\section*{Acknowledgements}   
We thank Mariarosaria D'Alfonso, Eva Halkiadakis, David Stuart, and Frank Wuerthwein  for discussions on the CMS Simplified Models. SC was supported in part by the Department of Energy under grant DE-FG02-13ER41986.  RE, JH, and ML were supported in part by the Department of Energy under grant DE-FG02-91ER40674.

\appendix{Appendix A: Relic Abundance and Direct Detection Numerics\label{sec:Appendix}}
 \label{appendix:numerical}
 The relic abundance is given approximately by
 \begin{eqnarray}
 \Omega_\chi h^2 & \simeq& 1.07\cdot 10^9\,  \mbox{GeV}^{-1} 
 \frac{x_f}{M_{\text{Planck}}g_{*S}^{1/2}(a+ 3b / x_f)}\\
 e^{x_f} &=& c(c+2)\sqrt{\frac{45}{8}}\frac{m_\chi M_{\text{Planck}}(a+ 6b / x_f)}{\pi^3 g_{*S}^{1/2} x_f^{1/2}}
 \end{eqnarray}
where $x_f = m_\chi/T_f \sim 25$ is the inverse freeze-out temperature and $g_{*S}$ is the relativistic degrees of freedom. In Appendix B, we list the formulas for $a, b$ for the models being considered.  
We use the values 
$c=\frac{1}{2}$ \cite{Kolb:1990vq}, $g_{*S}=100$, 
$\Omega_{\chi}h^2 = 0.1199 \pm 0.0027$ \cite{Ade:2013lta}.  
To calculate the direct detection scattering cross section, we use the 
matrix element values in Table~\ref{table:ddnumbers},
where the numbers (and notation) are taken from \cite{Hisano}.  

\begin{table}
  \caption{ Parameters for quark and gluon matrix elements \label{table:ddnumbers}} 
\label{table1}
\begin{center}
\begin{tabular}{ccc}
\begin{minipage}{0.25\hsize}
\begin{center}
\begin{tabular}{|l|l|}
\hline
\multicolumn{2}{|c|}{For proton}\cr
\hline
$f_{Tu}$& 0.023\cr
$f_{Td}$& 0.032\cr
$f_{Ts}$&0.020\cr
$f_{TG}$& 0.925\cr
\hline
\multicolumn{2}{|c|}{For neutron}\cr
\hline
$f_{Tu}$&0.017\cr
$f_{Td}$& 0.041\cr
$f_{Ts}$& 0.020 \cr
$f_{TG}$& 0.922 \cr
\hline
\end{tabular}
\end{center}
\end{minipage}
\begin{minipage}{0.5\hsize}
\begin{center}
\begin{tabular}{|l|l||l|l|}
\hline
\multicolumn{4}{|c|}{Second moment at $\mu=m_Z$ }\cr
\multicolumn{4}{|r|}{(for proton) }\cr
\hline
$G(2)$&0.48&&\cr
$u(2)$&0.22&$\bar{u}(2)$& 0.034\cr
$d(2)$&0.11&$\bar{d}(2)$&0.036\cr
$s(2)$&0.026&$\bar{s}(2)$&0.026\cr
$c(2)$&0.019&$\bar{c}(2)$&0.019\cr
$b(2)$&0.012&$\bar{b}(2)$&0.012\cr
\hline
\end{tabular}
\end{center}
\end{minipage}
\end{tabular}
\end{center}

\end{table}

\appendix{Appendix B: Model Details}
In this appendix, we collect our calculations for the annihilation   and direct detection cross sections for the different models.  We describe the Majorana dark matter model in detail and summarize the results for the other models.   

\subsection{Majorana Dark Matter}
In this model, the dark matter particle is a Majorana fermion and the quark partners are scalars. The Lagrangian for the new physics in two component notation is 
\beq
\mathcal{L} = |D_\mu Q|^2 -m_Q^2|Q|^2 +i \chi^\dagger \bar{\si} \cdot \partial \chi - \frac{1}{2}m_\chi(\chi^2+ \chi^{\dagger\, 2}) +\la (\chi q) Q^*+\la^* (\chi^\dag q^\dag) Q 
\eeq

\subsubsection{Relic Density}
The relic density is determined by the velocity-averaged annihilation cross section $ \left\langle \si v \right\rangle$ which is commonly parametrized by the coefficients 
\[  \left\langle \si v \right\rangle \simeq a+b v^2 \]
In this model, for the annihilation cross section $\chi \chi^\dag \rightarrow q q^\dag $, $a$ and $b$ are found to be 
\beq
a = \frac{3 m_\chi^2 \sqrt{1-r} r \la^4}{32 \pi  \left(m_Q^2-m_\chi^2 (r-1)\right)^2}, 
\eeq
\beq
  \begin{split}
b =& \frac{ \la^4 m_\chi^2}{256 \pi  \left(m_Q^2-m_\chi^2 (r-1)\right)^4 \sqrt{1-r}}
\big[-2 m_Q^2 m_\chi^2 r \left(22-35 r+13 r^2\right)\\
&+m_Q^4  \left(16-26 r+13 r^2\right)+m_\chi^4 (r-1)^2 \left(16-10 r+13 r^2\right) \big]
  \end{split}
\eeq
where $r \equiv m_q^2/m_\chi^2$. 
Thus, in the massless quark limit, the $s$-wave vanishes and we have the leading order results
\beq
a \overset{r \rightarrow 0}{\simeq} \frac{3 m_\chi^2 r \la^4}{32\pi(m_Q^2+m_\chi^2)^2}
\eeq

\beq
    b  \overset{r \rightarrow 0}{\simeq} \la^4\frac{m_\chi^2 \left(m_Q^4+m_\chi^4\right)}{16\pi \left(m_Q^2+m_\chi^2\right)^4  }
\eeq

\subsubsection{Direct Detection}
After integrating out the colored partner (since the typical momentum transfer at direct detection experiments is small compared to its mass) we get the low-energy effective Lagrangian\cite{Hisano-wino}, which is written in four component notation as

\beq \begin{split}
\mathcal{L}_\text{eff}= \sum\limits_q (f_q m_q \bar{\chi} \chi\bar{q}q+ d_q\bar{\chi}\gamma_\mu\gamma_5\chi\bar{q}\gamma^\mu\gamma_5 q+  \frac{g_q^{(1)}}{m_\chi} \bar{\chi}i\partial^\mu\gamma^\nu\chi\mathcal{O}^q_{\mu\nu}+\frac{g_q^{(2)}}{m_\chi^2}\bar{\chi}(i\partial^\mu)(i\partial^\nu)\chi\mathcal{O}^q_{\mu\nu} \\+f_G\bar{\chi}\chi G^a_{\mu\nu}G^{a\mu\nu} + \frac{g^{(1)}_G}{m_\chi}\bar{\chi}i\partial^\mu\gamma^\nu\chi\mathcal{O}^g_{\mu\nu}+\frac{g^{(2)}_G}{m_\chi^2}\bar{\chi}(i\partial^\mu)(i\partial_\nu)\chi\mathcal{O}^g_{\mu\nu})
\end{split} 
\eeq
where $\mathcal{O}^q_{\mu\nu}$ and $\mathcal{O}^g_{\mu\nu}$ are the twist-2 operators for quarks and gluons:
\beq\label{twist2q}
\mathcal{O}^q_{\mu\nu}\equiv\frac{1}{2}\bar{q}i\left(D_\mu\gamma_\nu+D_\nu\gamma_\mu-\frac{1}{2}g_{\mu\nu}\slashed{D}\right)q
\eeq
\beq\label{twist2g}
\mathcal{O}^g_{\mu\nu}\equiv\left(G_\mu^{a\rho}G^a_{\rho\nu}+\frac{1}{4}g_{\mu\nu}G^a_{\alpha\beta}G^{a\alpha\beta}\right)
\eeq
Then the spin-independent scattering cross section of DM with nucleons $(N=p,n)$ is obtained from the effective Lagrangian as
\beq\label{ddcross}
\si_{\chi N}=\frac{4}{\pi} \mu_N^2 |f_N|^2
\eeq
where $\mu_N$ is the reduced mass of the $\chi, N$ system.

The SI effective coupling $f_N$ is evaluated using the nucleon matrix elements of quark and gluon operators giving: 
\beq
\frac{f_N}{m_p}=\sum\limits_{q \,=\,u,d,s}f_qf_{Tq}+\sum\limits_{q \,=\,u,d,s,c,b}\frac{3}{4}(q(2)+\bar{q}(2))\left(g_q^{(1)}+g_q^{(2)}\right)-\frac{8\pi}{9\alpha_s}f_{TG}f_G 
\eeq
where
\begin{align}
& f_q=\frac{m_\chi\la^2}{16(m_Q^2-m_\chi^2)^2}  \\
 & g_q^{(1)}=\frac{m_\chi}{4}\frac{\la^2}{(m_Q^2-m_\chi^2)^2}
\\ &g_q^{(2)}=0 
\\ & f_G\simeq -\frac{\alpha_s m_\chi \la^2}{192\pi m_Q^2(m_Q^2-m_\chi^2)}\left(\sum\limits_{q \,=\,\text{all}} 1 +\sum\limits_{q \,=\,c,b,t} c_q \right) \label{eq:fGmajorana}
\end{align}
As explained in \cite{Hisano}, the first sum in $f_G$ extends over all quarks coupling to the dark matter and is the short distance contribution, whereas the second sum is the long-distance contribution of heavy quarks and has a  QCD correction factor $c_q = 1 + 11 \alpha_s(m_q)/4\pi$ \cite{Djouadi:2000ck}.  Following \cite{Hisano}, we take $(c_c, c_b, c_t) = (1.32,1.19,1)$.   The factors of $1/(m_Q^2-m_\chi^2)$ demonstrate the enhancement of the direct detection cross section when $Q, \chi$ become degenerate.  In particular, the twist-two terms proportional to $g_q^{(1)}$ are  strongly enhanced in this limit given the large values for $q(2)+\bar{q}(2)$.   
 
\FloatBarrier
\subsection{Dirac Dark Matter}

\subsubsection{Relic Density}

\beq
  a = \frac{3 m_\chi^2 \sqrt{1-r} \la^4}{32 \pi  \left(m_Q^2-m_\chi^2 (r-1)\right)^2}
\eeq
\beq
  \begin{split}
b =& \frac{ \la^4 m_\chi^2}{256 \pi  \left(m_Q^2-m_\chi^2 (r-1)\right)^4 \sqrt{1-r}}
\big[m_Q^4 \left(8-7 r+2 r^2\right)\\&+m_\chi^4 (r-1)^2 \left(-8+9 r+2 r^2\right)+2 m_Q^2 m_\chi^2 \left(-12+13 r+r^2-2 r^3\right) \big]
  \end{split}
\eeq
To lowest order in $r = m_q^2 / m_{\chi}^2$, 
\beq
a \overset{r \rightarrow 0}{\simeq} \frac{3 \la^4 m_\chi^2}{32\pi \left(m_Q^2+m_\chi^2\right)^2  }
\eeq
\beq
b \overset{r \rightarrow 0}{\simeq} -\frac{\la^4 m_\chi^2 \left(-m_Q^4+3 m_Q^2 m_\chi^2+m_\chi^4\right)}{32\pi \left(m_Q^2+m_\chi^2\right)^4  }
\eeq
so the cross section is not $s$-wave suppressed.

\subsubsection{Direct Detection}
We get a low energy effective Lagrangian
\beq 
\mathcal{L}_\text{eff} \sim \frac{\la^2}{8(m_\chi^2-m_Q^2)} (\bar{q}\gamma_\mu q\bar{\chi}\gamma^\mu\chi-\bar{q}\gamma_\mu\gamma^5q\bar{\chi}\gamma^\mu\gamma^5\chi) )
\eeq
The vector-vector interaction gives a spin-independent cross section, which is only dependent on interactions to the up and down quarks.  In our models, the coupling to up and down quarks is the same and following a few steps (see e.g. \cite{chacko10}) gives a cross section per nucleon
\beq
\si_{\chi N}=\frac{9 \la^4 \mu_N^2}{64\pi(m_\chi^2-m_Q^2)^2}. \label{DiracSigmaDD}
\eeq

\subsection{Real Scalar Dark Matter}
\subsubsection{Relic Density}
\beq
 a =\frac{3 m_\chi^2 (1-r)^{3/2} r \la^4}{4 \pi  \left(m_Q^2-m_\chi^2 (r-1)\right)^2} 
\eeq
\beq
b = \frac{m_\chi^2 \sqrt{1-r} r \left(9 m_Q^4 r+m_\chi^4 (r-1)^2 (-16+9 r)-2 m_Q^2 m_\chi^2 \left(16-25 r+9 r^2\right)\right) \la^4}{32 \pi  \left(m_Q^2-m_\chi^2 (r-1)\right)^4}
\eeq
Both $a$ and $b$ vanish as $r\rightarrow 0$.  
To lowest order,
\beq
 a \overset{r \rightarrow 0}{\simeq} r\frac{3 m_\chi^2 \la^4}{4\pi \left(m_Q^2+m_\chi^2\right)^2 } 
 \eeq
\beq
 b\overset{r \rightarrow 0}{\simeq} -r\frac{m_\chi^4 \left(2 m_Q^2+m_\chi^2\right) \la^4}{2\pi \left(m_Q^2+m_\chi^2\right)^4  }
 \eeq

\subsubsection{Direct Detection Cross Section}
The effective Lagrangian for SI scattering  for this model is
\beq
\mathcal{L}_\text{eff} = \sum\limits_q \left( f_q m_q \chi^2 \bar{q}q+\frac{g_q^{(1)}}{m_\chi^2} (\chi \partial^\mu \partial^\nu \chi) \mathcal{O}^q_{\mu\nu}\right)
\eeq
and 
\beq
 f_{q} =  \frac{\la^2}{2 (m_Q^2-m_\chi^2)}, \quad
 g_{q}^{(1)} =  \frac{\la^2 m_\chi^2}{ (m_Q^2-m_\chi^2)^2}.
 \eeq
In this Lagrangian, we have ignored terms with a $\gamma_5$ which are suppressed for nonrelativistic scattering.
We use the relationship between scalar and fermion matrix elements
\beq
 \frac{\langle\chi| \chi^2|\chi\rangle_\text{real scalar}}
 {\langle\chi| \bar{\chi}\chi| \chi\rangle_\text{Majorana fermion}} = \frac{1}{2m_\chi}
\eeq
where the denominator matrix element is twice as large compared to Dirac fermions, due to Majorana fermions being their own antiparticle. Thus, we can determine the spin-independent (SI) cross section by rescaling the Majorana case:
\beq
\si_{\chi N} =  \frac{\mu_N^2}{\pi}\left(\frac{f_N}{m_\chi}\right)^2
\eeq
where 
\beq
\frac{f_N}{m_p}= \sum\limits_{q \,=\,u,d,s}f_qf_{Tq}+\sum\limits_{q \,=\,u,d,s,c,b}\frac{3}{4}(q(2)+\bar{q}(2))g_q^{(1)} +\sum\limits_{q \,=\,c,b,t}\frac{2}{27}c_q f_q f_G 
\eeq
The second sum is the contribution from the twist-two operator and the third sum is the contribution from the heavy quarks $(c,b,t)$ to the nucleon mass \cite{Shifman:1978zn} and contains the QCD correction factor $c_q$.  Since we have not calculated the loop corrections to the gluons, we are unable to write down the short distance contributions to the scattering as in Eq.~(\ref{eq:fGmajorana}).  

\FloatBarrier

\subsection{Complex Scalar Dark Matter}
\subsubsection{Relic Density}
\beq
a = \frac{3 \la^4 m_\chi^2 (1-r)^{3/2} r}{16 \pi  \left(m_Q^2-m_\chi^2 (r-1)\right)^2}
\eeq
\beq
  \begin{split}
b =& \frac{\la^4 m_\chi^2 \sqrt{1-r}}{128 \pi  \left(m_Q^2-m_\chi^2 (r-1)\right)^4}
\Big[ m_\chi^4 (r-1)^2(9 r^2-18r+8) \\ &-2m_Q^2 m_\chi^2 (9r^3-31r^2+30 r-8) +m_Q^4 (9r^2-2r+8) \Big]
  \end{split}
\eeq
To lowest order in $r$ we get,
\beq
a \overset{r \rightarrow 0}{\simeq} \frac{3 \la^4 m_\chi^2 r}{16\pi (m_Q^2+m_\chi^2)^2} 
\eeq
\beq
b \overset{r \rightarrow 0}{\simeq} \frac{\la^4 m_\chi^2}{16\pi \left(m_Q^2+m_\chi^2\right)^2} 
\eeq
exhibiting the chiral suppression of $a$.  

The effective Lagrangian following \cite{chacko10} is
\beq
\mathcal{L}_\text{eff} = \frac{i \la^2}{2(m_Q^2-m_\chi^2)}\bar{q}\gamma^\mu q\, \chi^* \partial_\mu \chi
\eeq
where we have ignored terms suppressed in the non relativistic limit.  
For the complex scalar we can relate its direct detection scattering rate to the fermionic case via 
\beq
 \frac{\langle\chi| i \chi^* \partial^\mu\chi |\chi\rangle_\text{scalar}}
 {\langle\chi| \bar{\chi}\gamma^\mu\chi| \chi\rangle_\text{fermion}} = \frac{m_\chi}{2m_\chi}\delta ^\mu_0=\frac{1}{2}\delta ^\mu_0
\eeq

Rescaling from the Dirac dark matter cross section in  
Eq.~(\ref{DiracSigmaDD}), we find a complex scalar cross section 
\beq
\displaystyle \si_\text{SI} = \frac{9 \mu_N^2 \la^4}{16\pi (m_Q^2-m_\chi^2)^2}
\eeq

\subsection{Real Vector Dark Matter}
\subsubsection{Relic Density}
\beq
a = \frac{\la^4 m_\chi^2 \sqrt{1-r} (r^2-9 r + 8)}{12 \pi  \left(m_Q^2-m_\chi^2 (r-1)\right)^2}
\eeq
\beq
  \begin{split}
b =& -\frac{\la^4 m_\chi^2 \sqrt{1-r}}{288 \pi  \left(m_Q^2-m_\chi^2 (r-1)\right)^4}
\Big[m_\chi^4 (r-1)^2(17 r^2-92r+112) \\ &-2m_Q^2 m_\chi^2 (17r^3-69r^2+132 r-80) +m_Q^4 (17r^2-12r-80) \Big]
  \end{split}
\eeq
In the limit $r \rightarrow 0$,
\beq
a \overset{r \rightarrow 0}{\simeq} \frac{2 m_\chi^2 \la^4}{3\pi \left(m_Q^2+m_\chi^2\right)^2 }
\eeq
\beq
b \overset{r \rightarrow 0}{\simeq} -\frac{m_\chi^2 \left(-5 m_Q^4+10 m_Q^2 m_\chi^2+7 m_\chi^4\right) \la^4}{18\pi \left(m_Q^2+m_\chi^2\right)^4}
\eeq

\subsubsection{Direct Detection}
We get the following effective Lagrangian for spin-independent interactions \cite{Hisano}
\beq
\begin{split}
\mathcal{L}_q^\text{eff}=f^m_q m_q \chi^\mu\chi_\mu\bar{q}q
+\frac{g_q}{m_\chi^2}\chi^\rho i\partial^\mu i\partial^\nu\chi_\rho\mathcal{O}^q_{\mu\nu}
+f_G\chi^\rho\chi_\rho G^{a\mu\nu}G^a_{\mu\nu}
\end{split}
\eeq
where $\mathcal{O}^q_{\mu\nu}$ and $\mathcal{O}^g_{\mu\nu}$ are the twist-2 operators as in \eqref{twist2g} and \eqref{twist2q}.
The total spin-independent scattering cross section per nucleon is 
\beq\label{ddcross}
\si_{\chi N}=\frac{1}{\pi m_\chi^2} \mu_N^2 |f_N|^2
\eeq
where
\begin{align}
&\frac{f_N}{m_p}=\sum\limits_{q=u,d,s}f_qf_{Tq}+\sum\limits_{q=u,d,s,c,b}\frac{3}{4}(q(2)+\bar{q}(2)) g_q-\frac{8\pi}{9\alpha_s}f_{TG}f_G, \\
& f_q = -\frac{\la^2 m_Q^2}{4 \left(m_Q^2-m_\chi^2\right)^2},
\\ & g_q=-\frac{\la ^2 m_\chi^2}{\left(m_Q^2-m_\chi^2\right)^2},
\\ & f_G\simeq \frac{\alpha_s \la^2}{8\pi}\left(\sum\limits_{q=c,b,t}c_q \frac{m_Q^2}{6 \left(m_Q^2-m_\chi^2\right)^2}+\sum\limits_{q=all}
\frac{1}{3(m_Q^2-m_\chi^2)}\right).
\end{align}
Here, we have taken for simplification the limiting values of $f_{G}$ for the small $m_q$ limit.  Note that just like for the Majorana fermion model the first sum for $f_G$ is over the long distance contribution of the heavy quarks, where again there is a QCD correction factor $c_q$.

\FloatBarrier

\subsection{Complex Vector Dark Matter}
\subsubsection{Relic Density}
\beq
a = \frac{\la^4 m_\chi^2 \sqrt{1-r} (-r^2-7 r + 8)}{48 \pi  \left(m_Q^2-m_\chi^2 (r-1)\right)^2}
\eeq
\beq
  \begin{split}
b =& \frac{\la^4 m_\chi^2 \sqrt{1-r}}{1152 \pi  \left(m_Q^2-m_\chi^2 (r-1)\right)^4}
\Big[m_\chi^4 (r-1)^2(25 r^2-74r+40) \\ &+2m_Q^2 m_\chi^2 (-25r^3+167r^2-214 r+72) +m_Q^4 (25r^2-186r+296) \Big]
  \end{split}
\eeq
In the limit $r \rightarrow 0$,
\beq
a \overset{r \rightarrow 0}{\simeq} \frac{ m_\chi^2 \la^4}{6\pi \left(m_Q^2+m_\chi^2\right)^2 }
\eeq
\beq
b \overset{r \rightarrow 0}{\simeq} \frac{m_\chi^2 \left(37 m_Q^4+18 m_Q^2 m_\chi^2+5 m_\chi^4\right) \la^4}{144\pi \left(m_Q^2+m_\chi^2\right)^4}
\eeq

\subsubsection{Direct Detection}
The dominant vector-vector spin independent interaction is
\beq
\mathcal{L}_\text{eff} = -\frac{i \la^2}{2(m_Q^2-m_\chi^2)}\bar{q}\gamma^\mu q\, \chi_\nu^\dag \partial_\mu \chi^\nu.
\eeq
To relate its direct detection scattering rate to the fermionic case we use
\beq
 \frac{\langle\chi| i \chi_\nu^\dag \partial^\mu\chi^\nu |\chi\rangle_\text{vector}}
 {\langle\chi| \bar{\chi}\gamma^\mu\chi| \chi\rangle_\text{fermion}} = \frac{m_\chi}{2m_\chi}\delta ^\mu_0=\frac{1}{2}\delta ^\mu_0.
\eeq
Rescaling from the Dirac dark matter cross section in  
Eq.~(\ref{DiracSigmaDD}), we find a complex vector cross section 
\beq
\displaystyle \si_\text{SI} = \frac{9 \mu_N^2 \la^4}{16\pi (m_Q^2-m_\chi^2)^2}.
\eeq

\FloatBarrier

\bibliographystyle{utphys}
\bibliography{darkm}

\end{document}